\documentclass[9pt,twocolumn,twoside]{opticajnl}
\journal{opticajournal} 

\setboolean{shortarticle}{true}

\usepackage[]{changes} 
\newcommand{\add}{\textcolor{black}}
\newcommand{\remove}[1]{}
\usepackage{amsmath}
\usepackage{physics}
\usepackage{siunitx}
\usepackage{lineno}

\title{Polarization Tracking and Active Compensation Using Classical Headers in Quantum Wrapper Networking}

\author[1]{Gamze G\"{u}l}
\author[2]{James van Howe}
\author[1]{Gregory S. Kanter}
\author[1]{Shannon G. Tan}
\author[1,3]{Liam E. Beaudoin}
\author[4]{Mehmet Berkay On}
\author[5]{Roberto Proietti}
\author[4]{S. J. Ben Yoo}
\author[1,3,*]{Prem Kumar}

\affil[1]{Center for Photonic Communication and Computing, ECE Department, Northwestern University, Evanston, IL 60208, USA}
\affil[2]{Department of Physics, Engineering, and Astronomy, Augustana College, Rock Island, IL 61201, USA}
\affil[3]{Graduate Program in Applied Physics, Northwestern University, Evanston, IL 60208, USA}
\affil[4]{Department of Electrical and Computer Engineering, University of California, Davis, Davis, CA 95616, USA}
\affil[5]{Department of Electronics and Telecommunications Engineering, Politecnico di Torino, Italy}

\affil[*]{kumarp@northwestern.edu}

\begin{abstract}
 Quantum wrapper networking (QWN) is an emerging quantum networking protocol that wraps qubits in classical header bits to enable switching/routing, monitoring, and control without detecting the quantum signal. In this work, we encode header bits with two non-orthogonal polarization references to track and actively compensate for the changing birefringence of a 48 km deployed fiber link. Our method is analytical and deterministic, using motorized waveplates and a variable phase retarder to accurately and stably compensate the channel. We verify successful compensation by measuring the polarization stability of single photon qubits and the visibility of entangled photon pairs under both slow birefringence drift due to environmental fluctuations and large sudden changes designed to emulate those that occur during packet switching and rerouting over different fiber paths. \remove{For large, sudden changes, our compensator recovers the Stokes vector of single photons to within 10$^{\circ}$ of the target state on the Poincar\'{e} sphere and restores two-photon interference visibilities to better than 79\% on a deployed fiber link. Additionally, experiments monitoring long-term compensation over 44 hours show that visibilities remain above 84.5\% with compensation active and degrade to below the quantum threshold of 70.7\% within 4 hours of the compensator being turned off.}\add{For large, sudden changes, our compensator recovers the Stokes vector of single photons to within 10$^{\circ}$ of the target state on the Poincar\'{e} sphere and restores the two-photon interference visibilities to their original levels on the deployed fiber link. Additionally, experiments monitoring long-term compensation over 44 hours show that active polarization compensation maintains the two-photon interference visibility, whereas disabling the compensator leads to visibility degradation over time, particularly in the diagonal–antidiagonal basis, where the average raw visibility decreases by 7.7\%} These results add a polarization-control layer to QWN and illustrate that information-carrying headers can enable deterministic physical-layer compensation in the quantum channel over long-distance deployed fiber links. 
\end{abstract}

\setboolean{displaycopyright}{false} 

\begin{document}

\maketitle
\section{Introduction}

Quantum networks and quantum interconnects are emerging as enabling infrastructure for distributed quantum communication, sensing, and computing \cite{awschalom_development_2021,wehner_quantum_2018,kimble_quantum_2008,lee_quantum_2024, cacciapuoti_quantum_2020}. In numerous practical architectures, photonic qubits are generated at network nodes and transmitted through optical channels \cite{tanzilli_photonic_2005}. This approach establishes an effective interface between remote quantum platforms and facilitates entanglement distribution over long distances \cite{flamini_photonic_2018}. To this end, the last decade has seen an incredible growth in the implementation of quantum deployed fiber network testbeds \cite{liu_road_2025}. The last several years have seen impressive benchmarks obtained in entanglement distribution \cite{sena_high-fidelity_2025,rahmouni_100-km_2024,thomas_designing_2023, talcott_synchronized_2026}, teleportation \cite{thomas_quantum_2024, borowska_bichromatic_2026}, entanglement swapping \cite{sun_entanglement_2017,craddock_high-rate_2026}, quantum memories \cite{lei_quantum_2023,knaut_entanglement_2024}, long-distance Hong-Ou-Mandel (HOM) measurements \cite{ramesh_hong-ou-mandel_2025} and long-distance secret quantum key rates (SQKR) \cite{lu_device-independent_2026,chen_integrated_2021}. Notably, a number of these demonstrations have been performed in the presence of classical signals or classical data traffic, highlighting the feasibility of quantum–classical coexistence. Such coexistence will be essential for future quantum networks, which must operate alongside existing classical communication infrastructure and rely on classical control planes for network management \cite{chung_design_2022}. Despite this progress, the variation between methodologies and protocols remains large. Existing demonstrations differ widely in wavelength bands, entanglement types, multiplexing strategies, switching capabilities, classical control architectures, and requirements for technologies such as quantum frequency conversion or transduction. Similarly to classical communication, there needs to be an agreed upon protocol to control, monitor, and switch/route while qubits are transmitted within and between networks.

Quantum Wrapper Networking (QWN) \cite{ben_yoo_quantum_2024} is an emerging candidate to address this by “wrapping” quantum signals with classical headers and tails. These classical signals co-propagate with the quantum payload and encode, or allow inference of, critical information such as packet destination, entanglement type, payload duration, and channel phase or polarization drift. In the first demonstration of QWN, we showed successful routing and monitoring of QWN polarization entangled data in a three-node, laboratory fiber network testbed \cite{on_experimental_2024}. In another demonstration, we quantified the noise impact of classical QW headers on polarization entangled photons and showed that it had negligible impact on the entanglement distribution (measured locally) over 48 km of deployed fiber \cite{gul_noise_2025}. 

In this work, we leverage classical quantum wrapper  headers, already used for general network control and management, as polarization probes for quantum datagrams \cite{gul_active_2025}. Our system compensates both for slow drift due to environmental fluctuations on the link and sudden large changes in birefringence that we purposely introduce to emulate abrupt changes that qubits would experience, for instance, when being routed along different paths. Specifically, we prepare header bits with two non-orthogonal polarization states and measure their transformed polarization state to estimate the effective polarization rotation and relative phase retardance associated with the fiber transformation. The polarization evolution can be mapped to a rotation on the Poincar\'{e} sphere, enabling the direct computation of a compensating transformation \cite{xavier_full_2008}. We then actively apply the inferred correction with the calculated compensating rotations to stabilize the co-propagating quantum payload. Our compensator uses free-space motorized waveplates (quarter waveplate and half waveplate) and a free-space variable liquid crystal retarder and our method relies on a deterministic, analytical approach to calculate and apply the compensation. We tested our setup by monitoring the polarization state of single photon qubits and the visibility of entangled qubits transmitted over a 48 km deployed fiber link with their headers. Our experiments demonstrate that polarization compensation can maintain single photon polarization to within \ang{10} of a target input state on the Poincare sphere and maintain two-photon \add{interference (TPI)} visibilities \remove{above 79\%}\add{at their original levels} despite sudden large birefringence changes that we purposely introduce to emulate quantum packet rerouting over different fiber paths on a realistic network. Additionally, our experiments monitoring long-term compensation over 44 hours show that, after the compensation is disabled at hour 32, the average visibilities decrease by about \remove{1}\add{0.9}\% and \remove{12}\add{7.7}\% in the HV and DA bases, respectively. Our overall results suggest that active birefringence compensation can be seamlessly integrated into QWN, enabling a robust protocol for the transmission of quantum information in a realistic quantum network and paving a way for the eventual integration of different quantum networks with each other.

\section{Role of Active Polarization Control in QWN Architecture}\label{section:role}

Polarization encoding for qubits remains widely used because it supports straightforward state preparation and analysis with passive optics and can realize high-fidelity polarization entanglement and high brightness in deployable systems \cite{li_optical-fiber_2005,wengerowsky_entanglement-based_2018, kim_phase-stable_2006, wengerowsky_entanglement_2019, fedrizzi_wavelength-tunable_2007, treiber_fully_2009, du_entanglement_2025}. However, preserving polarization-encoded states over real-world links can be nontrivial \cite{ulrich_polarization_1979, martinelli_polarization_2006, yoshino_maintenance-free_2013}. For example, standard single-mode fibers exhibit spatially varying birefringence so that the state of polarization (SOP) undergoes a time-dependent transformation. This rotates polarization states, which reduces transmission and measurement performance such as entanglement visibility \cite{rashleigh_origins_1983, agrawal_chapter_2013-1}. Field measurements indicate that fiber link dynamics are highly dependent on deployment conditions. Aerial or partially aerial links may experience rapid fluctuations, whereas buried or submarine fibers tend to be more stable but still exhibit gradual drift over time  \cite{ding_polarization_2017,brodsky_polarization_2006,wengerowsky_entanglement_2019, amies-king_quantum_2023,du_entanglement_2025, ramesh_characterizing_2025}. Consequently, polarization stabilization has become a fundamental engineering requirement for polarization-encoded quantum key distribution (QKD), entanglement distribution, and other polarization-sensitive quantum networking tasks \cite{rubenok_real-world_2013, li_field_2018, yin_polarization_2025}. Existing approaches can be categorized as using passive or active strategies. Passive methods typically rely on pausing system operation to perform manual or automated realignment, thereby mitigating slow polarization drifts \cite{wengerowsky_passively_2020}. Active polarization control employs feedback to continuously compensate for channel fluctuations, either by directly optimizing a quantum performance metric (e.g., quantum bit-error-rate (QBER), count rates, or interference visibility) or by using a reference signal to track and correct polarization transformations in real time \cite{peranic_study_2023, kucera_demonstration_2024, shi_fibre_2021, xavier_full_2008, neumann_continuous_2022,treiber_fully_2009,stevens_situ_2025,tan_real-time_2024}. In many deployed demonstrations, active control is enabled by co-propagating classical references in time and/or as wavelength-multiplexed signals to estimate polarization transformations without relying on quantum counts \cite{xavier_experimental_2009, chapman_continuous_2024,craddock_automated_2024,zhang_classical-decisive_2025}. In fact, recent systems emphasize minimizing "downtime" as a performance metric, requiring little to no use of qubits as system controls (recognized as a critical resource, qubits are to be preserved for quantum information). For example, continuous stabilization via heterodyne detection of coexisting dim reference signals has been demonstrated to maintain high-fidelity operation without interrupting the quantum channel \cite{chapman_continuous_2024}. Independently, long-duration field operation with automated compensation has been shown in metropolitan fiber environments, illustrating that robust polarization stabilization can support a nearly continuous entanglement distribution \cite{craddock_automated_2024, shi_entanglement_2026}. More recently, classical-decisive quantum networking architectures based on integrated photonics have demonstrated mitigation of channel-induced errors using co-propagating classical header signals over short fiber links ($\sim$1\,km), with error mitigation performed at a server node and enabled by two-way communication \cite{zhang_classical-decisive_2025}.

\begin{figure*}[h!]
\centering\includegraphics[width=0.9\linewidth]{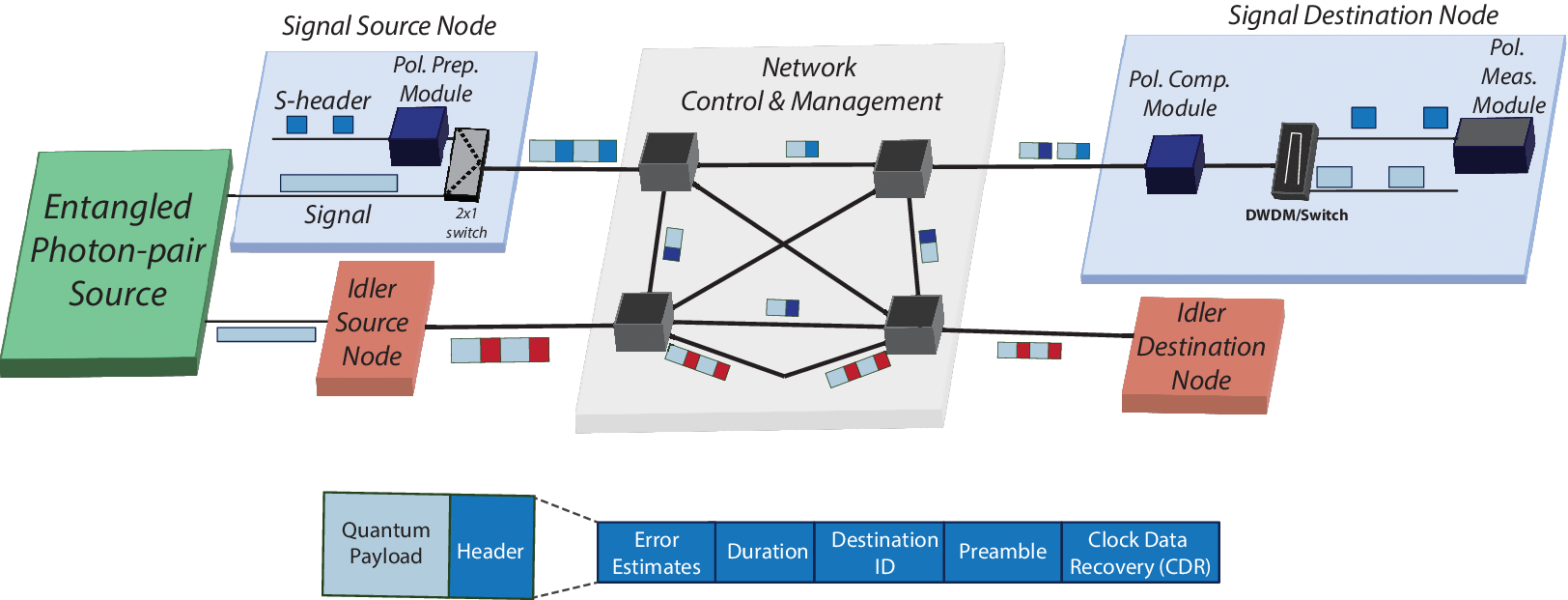}
\caption{Integration of active polarization control modules in QWN for the distribution of entangled photon pairs in the network. Classical headers are created in the source nodes to carry two non-orthogonal polarizations in addition to the information bits as shown at the bottom. Signal and idler photons are transmitted through the network after they are wrapped with these headers at the source nodes. The datagrams can travel in different paths as shown in signal datagrams or in a single route as in the case of idler datagrams. After the polarization states are measured at the destination nodes, polarization compensation modules at each node can compensate for the changes due to the drifts in these fiber links as well as the various transfer matrices due to different routes. \label{intro_fig}}
\end{figure*}

In QWN, quantum information travels along with its classical information bits, which are inserted before and/or after the quantum payload as a header and/or tail, respectively. In addition to carrying metadata about the quantum state, the measurements performed on the headers can reveal link properties experienced by the payload, including timing and polarization transformations. In the type of reconfigurable network setting we envision, where a packet may traverse a variety of different paths, dynamic measurement and control of timing and polarization information will be essential. Figure~\ref{intro_fig} depicts the transmission of a quantum payload in a packet-switched network. The packetized quantum information can travel along multiple possible routes, each with distinct polarization transfer functions that are also experienced by their corresponding headers. Although the paths traveled by the photons may differ, the QWN must guarantee the tracking of the corresponding polarization transformation for each channel via measurements performed on the headers. A polarization state compensation module deployed at destination node can subsequently compensate for polarization transformations based on the header polarization state and its arrival time.

Additionally, in an entanglement distribution where both signal and idler photons travel to distant nodes, an optimization method based on a quantum metric like fidelity might be overly time consuming, particularly as link loss increases and quantum statistics become sparse \cite{neumann_continuous_2022}. In those cases, each photon may need its own reference signals to copropagate. Although this is a feasible solution for small and/or static systems, it might limit the bandwidth of the transmissions in multi-user networks as the channels get noisier with increasing number of reference signals. In QWN, the entanglement fidelity is maintained by independently compensating polarization changes in both signal and idler links by taking the polarization measurements on the headers. If it is performed during the header timespan, this guarantees the entanglement distribution quality without interruption. In this work, we test the possibility of using headers for polarization compensation without optimizing for the speed.  

\section{Polarization Stabilization Method: Theory and Operational Principle}\label{section:theory}
\subsection{Compensation Method} \label{subsection:comp_method}
\begin{figure}[htbp] 
\centering\includegraphics[width=0.9\linewidth]{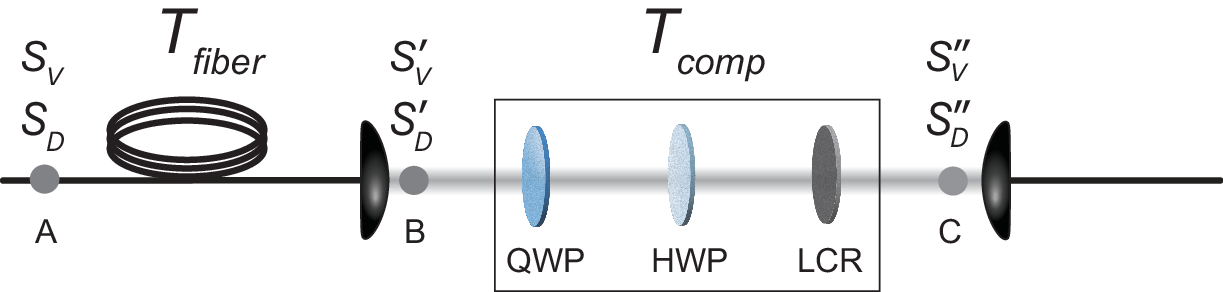}
\caption{Schematic of the polarization compensation architecture. 
The fiber transfer function $T_{\mathrm{fiber}}$ transforms the input reference states 
$S_V$ and $S_D$ at node A into $S'_V$ and $S'_D$ at node B. 
A compensation module implementing $T_{\mathrm{comp}}$, composed of a quarter-wave plate (QWP), 
half-wave plate (HWP), and liquid crystal retarder (LCR), is then used to recover the target 
polarization states $S''_V$ and $S''_D$ at node C.}
\label{fig:theory_fig}
\end{figure}

We implement an analytical, deterministic birefringence-compensation technique that infers the inverse polarization transfer matrix of the fiber link from two non-orthogonal reference states. Denoting the fiber polarization transfer matrix as $T_{\mathrm{fiber}}$ and the compensator as $T_{\mathrm{comp}}$, ideal compensation corresponds to
$T_{\mathrm{comp}} = T_{\mathrm{fiber}}^{-1},$
so that any incident polarization state $S_{\mathrm{in}}$ is restored at the output,
$S_{\mathrm{out}} = T_{\mathrm{comp}} \, T_{\mathrm{fiber}} \, S_{\mathrm{in}} = S_{\mathrm{in}}$. Figure~\ref{fig:theory_fig} schematically shows the fiber transformation (A$\rightarrow$B) followed by the compensator (B$\rightarrow$C).

We apply $T_{\mathrm{comp}}$ through two typical rotations used to arbitrarily control single polarization states: a quarter waveplate (QWP) and half waveplate (HWP), and a third- a phase retarder (liquid crystal retarder, LCR), to fully compensate the birefringence of the channel,
\begin{equation}
T_{\mathrm{comp}}=R_{\mathrm{LCR}}{R_{\mathrm{HWP}}R}_{\mathrm{QWP}}.	
\label{rot_gamze}
\end{equation}

Though any pair of non-orthogonal states can be used as inputs to determine $T_{\mathrm{comp}}$, deliberately choosing conjugate states (bases) maximizes the accuracy of reproducing the correct transfer function to fully compensate the channel. In our demonstration, we use conjugate pairs (V), vertical polarization, and (D), diagonal polarization such that the fiber link transforms linear input reference states $S_V$ (V) and $S_D$ (D) to arbitrary elliptical states $S_V^\prime$ and $S_D^\prime$ from point A to point B in Fig.~\ref{fig:theory_fig} as
\begin{equation}
\centering
\begin{split}
     S_V^\prime&= T_{\mathrm{fiber}}S_V \\
     S_D^\prime&= T_{\mathrm{fiber}}S_D.
\end{split}
\label{stokes_gamze}
\end{equation}

\begin{figure}[htbp] 
\centering\includegraphics[width=0.9\linewidth]{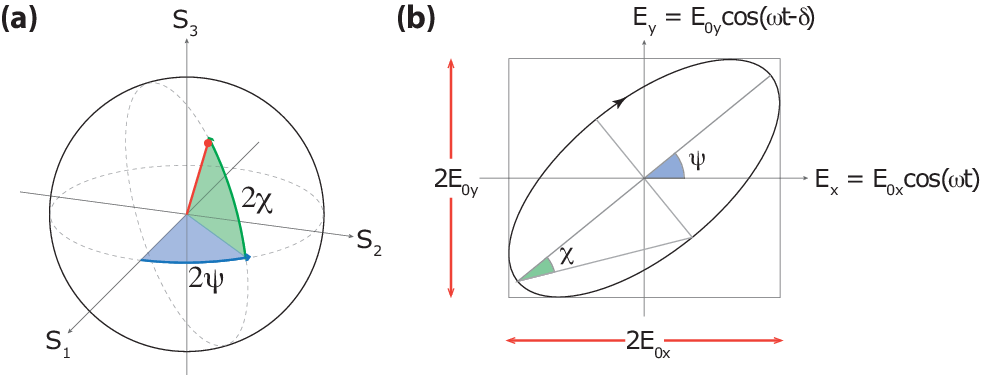}
\caption{\add{Connections between (a) Poincaré sphere and (b) polarization ellipse formalisms}}
\label{fig:stokes_jones}
\end{figure}

An illustrative description can be given using the Poincar\'{e} sphere representation together with the polarization ellipse formalism, \add{Fig. \ref{fig:stokes_jones}}. The angles $\psi$ and $\chi$ describe the polarization state in both the Stokes and Jones representations \cite{damask_properties_2005, agrawal_chapter_2013-1}. 

Specifically, $2\psi$ denotes the azimuth angle of the Stokes vector on the Poincar\'{e} sphere, while $\psi$ corresponds to the orientation angle of the semi-major axis of the polarization ellipse. Similarly, $2\chi$ represents the inclination (latitude) angle of the Stokes vector on the Poincar\'{e} sphere, whereas $\chi$ defines the ellipticity angle of the polarization ellipse. The phase difference $\delta$ between the $x$- and $y$-components of the electric field in the Jones representation is related to the Stokes parameters and corresponds to the inclination angle when the Stokes vector is projected onto the $S_2$--$S_3$ plane.

Assuming normalized Stokes parameters, these quantities can be calculated as
\begin{equation}
2\psi = \arctan\!\left(\frac{S_2}{S_1}\right),\;
2\chi = \arcsin(S_3), \;
\delta = \arctan\!\left(\frac{S_3}{S_2}\right).
\end{equation}

The appropriate rotation angles for the QWP, HWP, and the phase applied by the LCR are obtained by determining the transformations that simultaneously restore $S_V^{\prime}$ and $S_D^{\prime}$ to their respective reference states $S_V = V$ and $S_D = D$.
                                   
These are found as follows:
\begin{enumerate}

    \item 	Aligning the QWP (fast axis) with the semi-major axis of the ellipse of $S_V^\prime$, $\theta_{\mathrm{QWP}}$= $\psi_{S_V^\prime}$, which projects $S_V^\prime$ onto the $S_1$-$S_2$ plane (brings the elliptical state to linear polarization)
    
    \item 	Aligning the HWP (fast axis) to the half-angle of the linear state from 1) with the desired polarization state, (V), $\theta_{\mathrm{HWP}}=\left(\psi_{S_V^\prime}-\chi_{S_V^\prime}\right)/2-\frac{\pi}{4}$
    
    \item Trimming the phase of the rotated state $S_D^{R\prime}=R_{\mathrm{HWP}}R_{\mathrm{QWP}}S_D^\prime$, which is projected onto the $S_2$-$S_3$ plane by the rotations from the QWP and HWP, where the phase is determined as $\Gamma_{\mathrm{LCR}}=-\delta_D^{R^\prime}$ or $-2\chi_D^{R\prime}$
\end{enumerate}

Here rotations 1) and 2) are similar to those typically used to convert elliptical polarization states to linear states. Because the angles in our equations are defined with respect to the x-axis, the $\pi/4$ term in 2) is required to return $S_V^\prime$ to (V) for our system, else it would be (H). 

In our setup, a polarimeter periodically measures the Stokes parameters of the output polarization states of the system at point C, (Fig.~\ref{fig:theory_fig}). If $S_V^{\prime\prime}$,$S_D^{\prime\prime}=S_V,S_D$, then $T_{\mathrm{comp}}T_{\mathrm{fiber}}=1$ and no compensation is applied. However, when $S_V^{\prime\prime},S_D^{\prime\prime}\neq S_V,S_D$ (within a user-specified tolerance), a polarization compensation cycle is applied. First, the states at point B are calculated from the measured polarimeter values as $S_V^\prime, S_D^\prime=T_{\mathrm{comp}}^{-1}S_H^{\prime\prime},S_D^{\prime\prime}$ , where $T_{\mathrm{comp}}^{-1}$ is the inverse of the current state of the compensator. Next, the appropriate rotations of the QWP, HWP, and the LCR are analytically determined using steps 1)-3) above and applied by motorized rotation stages for the QWP and HWP, and as the appropriate LCR voltage for the required phase. This process is iterated until $S_V^{\prime\prime}$ and $S_D^{\prime\prime}$ match $S_V$ and $S_D$ within the user-specified tolerance. In principle, a single iteration is sufficient; however, in practice, multiple iterations may be required due to measurement noise, calibration errors, and resolution limits (see Sec.~\ref{section:results}\ref{subsection:comp_time} for detailed analysis).

\subsection{Effects from Polarization Mode Dispersion}\label{subsection:pmd_effects}

Due to polarization mode dispersion (PMD), which causes wavelength-dependent polarization rotation (first order PMD) and wavelength-dependent birefringence (second order PMD), perfect birefringence channel compensation can only happen if the signal we wish to compensate (here, the quantum payload) is at the same wavelength as the polarization reference headers, since $T_{\mathrm{fiber}}\left(\omega_{\mathrm{pay}}\right)T_{\mathrm{comp}}\left(\omega_{\mathrm{head}}\right)=1$ when $\Delta\omega=\omega_{\mathrm{header}}-\omega_{\mathrm{payload}}=0$. However, polarization compensation remains approximately valid, $T_{\mathrm{fiber}}\left(\omega_{\mathrm{pay}}\right)T_{\mathrm{comp}}\left(\omega_{\mathrm{head}}\right)\approx1$  if $\Delta\omega\Delta\tau \ll 1$, where $\Delta\tau$ is the mean differential group delay (DGD) of the fiber due to PMD (“monochromatic limit”) \cite{xavier_full_2008}. Because QWN leverages the noise robustness and flexibility of wavelength-division multiplexing (also critical for current and planned fiber networks), we are specifically interested in $\Delta\omega \neq 0$ and investigate the largest channel separation we can achieve for the fiber distances and PMD parameters of the fibers employed.

In our setup, we find that keeping $\Delta\omega\Delta\tau < 0.42$ \cite{howe_polarization_2025} meets the "monochromatic" limit, corresponding to a maximum channel separation $\Delta\lambda = 0.7$\,nm (122 GHz at 1310 nm) for the fibers used.
This limitation can be more easily understood by looking into how first-order PMD results in wavelength-dependent polarization rotation. Given a small enough bandwidth where second-order PMD is negligible, $\Delta\omega\Delta\tau < \pi/4$ \cite{kogelnik_polarization-mode_2002, damask_properties_2005}, the fiber link can be modeled as a non-mode coupled device dominated by first-order PMD. In this case, the same input polarization states to a fiber at different wavelengths will transform to different output polarization states on the Poincar\'{e} sphere. In this limit, the output polarization states will rotate through an angle $\Delta\theta$ on the sphere as a function of wavelength about the PMD vector axis. 

A non-zero arc length between these vectors will lead to inaccurate determination of the fiber compensation matrix since we require $S_V^\prime{(\omega}_{\mathrm{head}})=S_V^\prime{(\omega}_{\mathrm{pay}})$ and $S_D^\prime{(\omega}_{\mathrm{head}})=S_D^\prime{(\omega}_{\mathrm{pay}})$ for perfect compensation. 
Therefore, the arc length, given by $\ell_{\Delta\omega}=\Delta\omega\Delta\tau \sin(\varphi)$, where $\varphi$ is the angle of the cone swept out by Stokes vectors at different channel wavelengths, is a measure of the error. When the input polarization to the fiber is aligned along one of the principal states of polarization (PSP) vectors of the fiber, $\varphi=0$,  $\ell_{\Delta\omega}= 0$, we can ideally use headers to perfectly determine the fiber compensation matrix since the arc length is zero. However, in the worst-case scenario where the input polarization is aligned orthogonally to the PSPs, $\varphi=\ang{90}$, $\ell_{\Delta\omega}=\Delta\omega\Delta\tau $. Constraining the system to make this as small as possible, $\Delta\omega\Delta\tau \ll 1$, allows effective compensation, albeit approximate. Because  $\Delta\tau = \mathrm{PMD}\,\sqrt{L}$, where L is the length of the fiber, and "PMD" in the equation is the PMD parameter (typically measured in $\sqrt{{\mathrm{ps}}\cdot{\mathrm{km}}}$), one typically limits the channel separation, $\Delta\omega$, in favor of transmission distance. However, when applicable, large channel separations can be employed for shorter fiber links and/or fibers with low PMD parameters. 

Although aligning the input polarization states with the PSPs can, in principle, mitigate first-order PMD, our system is designed to compensate birefringence under dynamically changing fiber paths (e.g., different routing configurations of quantum payloads across network nodes). Therefore, we cannot assume prior knowledge of the fiber PSPs and must design the compensation scheme to operate under worst-case conditions. However, we note that several recent works have investigated PMD mitigation strategies in polarization compensation schemes for quantum networking, particularly in scenarios where spectral separation between reference and quantum channels is required \cite{xavier_full_2008,howe_polarization_2025,kogelnik_polarization-mode_2002,eastman_direct_2025, stevens_situ_2025}.

\section{Experimental Setup} \label{section:setup}
The experimental setup, Figs.~\ref{fig_pol_prep}-\ref{fig:pol_comp}, integrates polarization tracking and active birefringence compensation as an additional level of control to the QWN system demonstrated in \cite{gul_noise_2025}. Our system can be divided into three submodules:\,1) Header Polarization State Preparation, 2) Polarization State Measurement, 3) Polarization Stabilization, each of which we discuss in detail below. We additionally show the baseline performance/calibration of the system with no fiber present and describe the deployed fiber link used in this work.

\subsection{Polarization State Preparation Module}\label{subsection:prep}

\begin{figure}[h!] 
\centering\includegraphics[width=\linewidth]{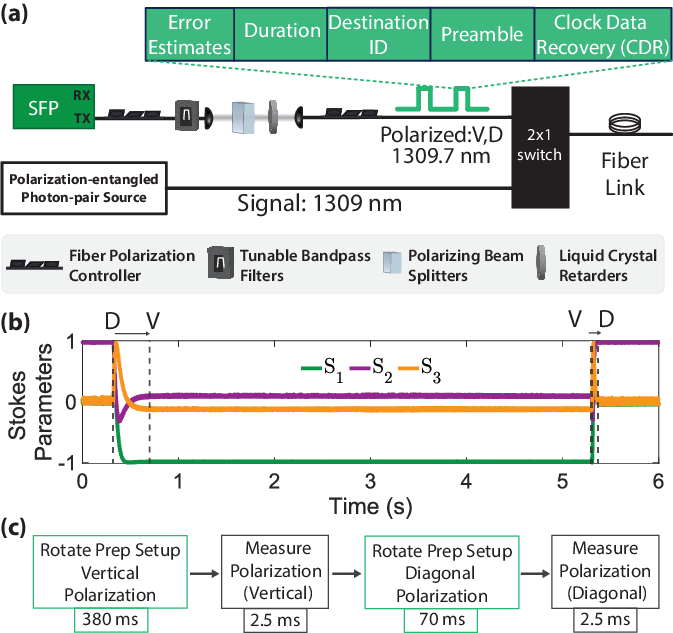}
\caption{a) Experimental setup to create quantum wrapper (QW) headers with two non-orthogonal polarizations. The headers with vertical and diagonal polarizations are created sequentially and quantum payload is multiplexed with the headers by using a fast switch. b) Measurement of the switching time of the LCRs to create vertical and diagonal states. These transitions are observed by altering the RMS drive voltage of the LCR every 5\,s. The transition from D to V is captured between 0-1\,s and V to D between 5-6\,s. c) The order of tasks to capture transfer matrix of the fiber link. Green boxes correspond to the waiting time we use in our method. Each polarization preparation cycle is followed up with the measurement of the header polarization at the destination. }
\label{fig_pol_prep}
\end{figure}

The function of the polarization state preparation module, Fig.~\ref{fig_pol_prep}, is to encode non-orthogonal polarization references on QWN headers, which then probe the fiber under test (see Sec.~\ref{section:theory} above). We choose polarization states (V) vertical and (D) diagonal as references, which are imposed on the classical reference headers by passing the headers through a polarization beam splitting cube (PBS) and a variable liquid crystal retarder (LCR), whose phase is toggled after a 2.5\,ms sampling period, Fig.~\ref{fig_pol_prep}(c). Here, the PBS fixes the polarization state to (V), the fast axis of the liquid crystal retarder is set to \ang{67.5}, and the LCR phase is varied from 0 to $\pi$. In this way, the input vertical polarization state rotates about the axis that splits the $-S_1$ and $S_2$ axes (2$\times$\ang{67.5} = \ang{135}) when the LCR phase is varied. When the LCR phase equals 0, the state remains in (V), aligned along the $-S_1$ axis. When toggled to $\pi$, the state rotates to the $S_2$ axis to (D).

We use full-wave LCRs (Thorlabs, LC1513-C, and their corresponding drivers). The LCR switching time is on the order of a few hundreds of milliseconds, as shown in Fig.~\ref{fig_pol_prep}(b). The drivers provide a square wave with a frequency of 2\,kHz and a range of root-mean-square (RMS) voltages of 0-25\,V.  We determine the voltage values to create vertical and diagonal polarizations (LCR voltage for a phase of 0 and $\pi$, respectively) by characterizing the LCRs in advance.  

Headers are generated using a 1\,Gbps small form-factor pluggable (SFP) transceiver from Solid Optics as described in \cite{on_experimental_2024}. The header has a duration of 345.8\,$\mu$s, a repetition rate of $\sim$604\,Hz, and a wavelength of 1309.7\,nm.  We use a bandpass filter and an FPC before the polarization preparation setup to suppress the side bands of the SFP at our quantum payload wavelength (1309\,nm) and align the polarization of the SFP to the PBS of the polarization preparation setup. 

Since the switching time ($<$380\,ms) is much longer than the header duration (345.8\,$\mu$s), we only measure the polarization of the header after waiting for the LCR switching time. After creating the polarization states, we measure headers one by one and measure the polarization state at the destination node for 2.5\,ms before switching to the second reference. Fig.~\ref{fig_pol_prep}(c) shows the protocol and how the switching time and header measurement duration constrain polarization state preparation.  

\subsection{Polarization State Measurement Module}\label{subsection:measure}

\begin{figure}[h!]
  \centering
  \includegraphics[width=\linewidth]{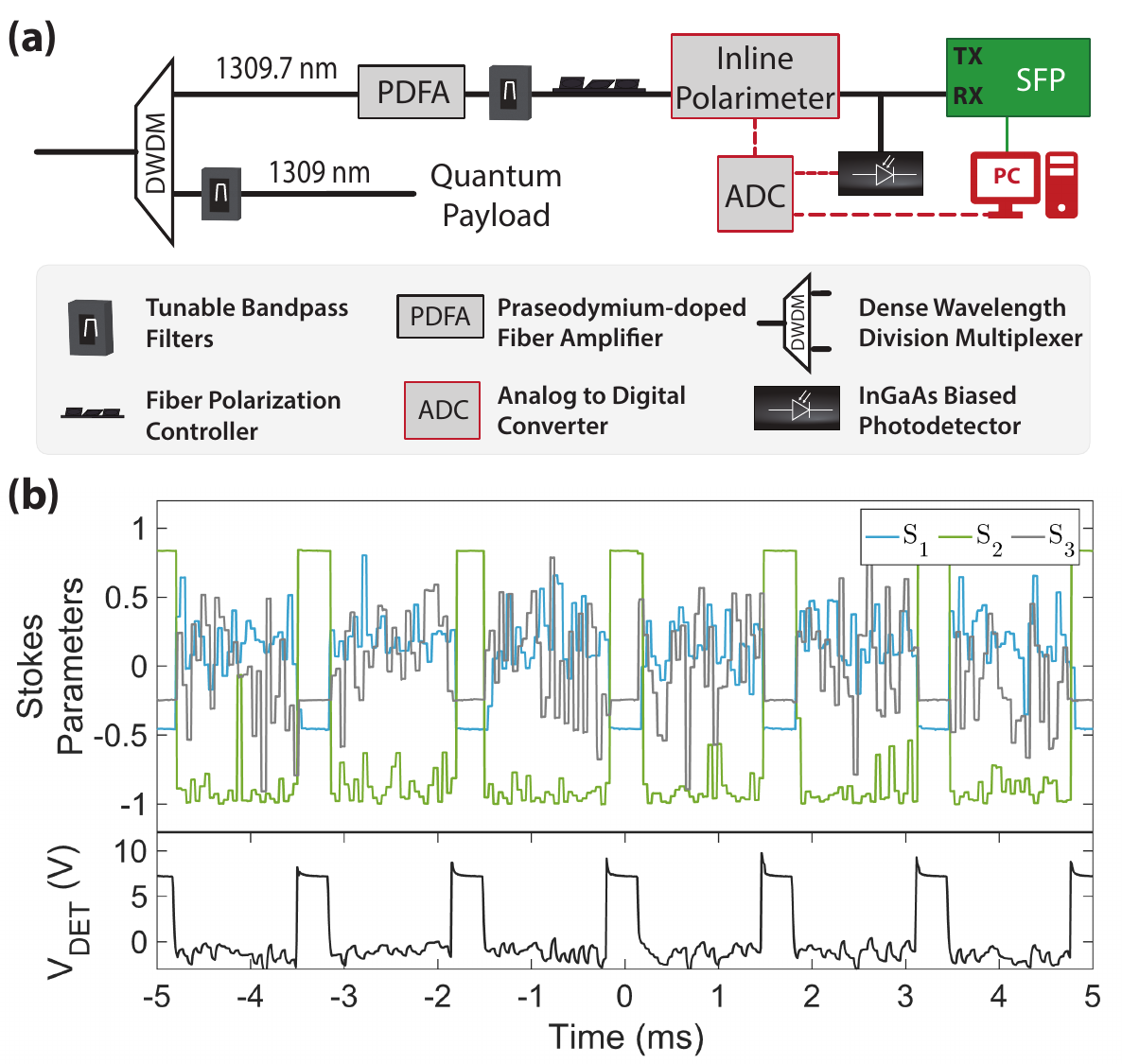}
  \caption{a) The header polarization measurement setup. Quantum payloads and headers are demultiplexed and headers are amplified by using a polarization insensitive amplifier. The inline polarimeter outputs three stokes parameters. b) The ADC also takes a synchronous classical signal in order to gate the measurements during the header time.} 
  \label{fig:pol_meas} 
\end{figure}

In QWN, the destination node is equipped with a polarimeter to measure the polarization state of the headers. In our experiments, we use a TXP inline polarimeter from Thorlabs with a sampling time of 10\,$\mu$s\add{, as shown in Fig.~\ref{fig:pol_meas}(a)}. 
This provides Stokes parameters $S_1$, $S_2$, and $S_3$ from values ($-$1,1) as voltages between $-$2.5 to 2.5\,V. We take voltage samples using an analog to digital converter on a data acquisition (DAQ) device (MCC DAQ 1208HS). We additionally trigger the DAQ with the detected optical output from the polarimeter (the TPX uses 10\% of the input signal for detection of Stokes parameters and passes the remaining to the output). We tap off 99\% from the polarimeter's output and detect it with a biased photodiode detector (Thorlabs DET01CFC). This method allows us to effectively gate the polarization measurements during the short header duration and ignore random measurements happening during the quantum payload's timespan.

The sensitivity of the polarimeter is also important for the accurate measurement of the Stokes vectors. In our experiments, we use a praseodymium-doped fiber amplifier (PDFA) with a 24\,dB signal gain and a minimum input power of $-$40\,dBm to amplify our headers. We also use a tunable bandpass filter in order to suppress amplified spontaneous emission (ASE) \add{generated by the PDFA} before the polarization measurement. This process allows us to measure the polarization state accurately while maintaining their polarization states.

\subsection{Active Polarization Compensation Module}\label{subsection:comp}

Our experimental setup for the polarization compensation uses free-space components: a QWP, HWP, and an LCR. The waveplates are mounted on Thorlabs motorized rotation stages, and the LCR is controlled by varying voltage with a Thorlabs KLC module. Faster components can enable higher-speed compensation, such as using three LCRs for the required rotations (switching speed of $\sim$500\,ms) or electro-optic polarization controllers (speed of $\sim$100\,ns) \cite{chapman_continuous_2024}. However, for this first demonstration of QWN birefringence compensation, we choose relatively slower free-space components ($\sim$10\,s) because of their simplicity, stability, and accuracy. Indeed, future iterations of our system will include increasing compensation cycle speed using different optical components in the compensator.   

A birefringence compensation cycle occurs if $S_V^{\prime}$, deviates from its ideal input state V Stokes parameter $S_1$ value by 0.1\% ($<$0.999) or if $S_D^{\prime}$ deviates from its ideal input state D Stokes parameter $S_3$ value by 0.5\% ($<$0.995). These thresholds can be chosen independently to reflect different experimental sensitivities. In our experiments, the diagonal state is more susceptible to calibration and preparation errors, and therefore a slightly relaxed threshold is used to provide sufficient margin for experimental uncertainty while maintaining overall performance. In the compensation cycle, the correct compensating angles on the QWP, HWP, and compensating phase from the LCR are calculated from the Stokes parameters of $S_V^{\prime}$ and $S_D^{\prime}$ (see Sec.~\ref{section:theory} for details) and applied accordingly.

\begin{figure}[h!]
  \centering
  \includegraphics[width=\linewidth]{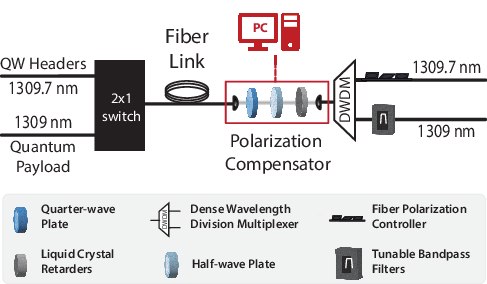}
  \caption{Polarization compensation and channel stabilization module based on free space HWP, QWP, and LCR. The compensation module is activated if the polarization measurements demonstrate a drift more than the threshold set initially.} 
  \label{fig:pol_comp} 
\end{figure}
Both waveplates are mounted on PRMZ01 motorized rotation stages from Thorlabs. In our setup, they rotate at a speed of 15\,deg/s. However, the compensation speed is a function of the angles of the required rotation, which changes based on how large an angle needs to be applied. To illustrate, we can take a rerouting scenario that rotates the polarization from R to L. The compensator then rotates the QWP by \ang{180}, which takes 12\,s at the angular speed noted above. In our particular setup, we use a USB hub for all rotation stages that rotates one waveplate at a time. Therefore, the compensation time is a function of the sum of the rotation angles. The shortest compensation time in our setup is 4.8\,s and the largest is 98.6\,s (see Sec.~\ref{section:results}\ref{subsection:comp_time} below for more details). The LCR switching time, $\sim$400\,ms for the slowest transition, is not a limiting factor in the compensation cycle speed.

Compensation of phase using the LCR is performed by finding the voltage that corresponds to the required retardance $\delta$ using a look-up table in our code (made by characterizing the LCR beforehand).The voltage that adds the phase retardance closest to the one found by calculation is selected. Note that the resolution of the voltage measurement representing the phase and the temperature dependence of the phase retardance function are potential sources of error \cite{schnoor_calibration_2020}. However, the former is easily solved by taking higher resolution calibration curves while the latter may require environmental isolation or temperature stabilization. 

Finally, we note that the compensator only compensates the birefringence between the 2$\times$1 switch and the DWDM in Fig.~\ref{fig:pol_comp}. The fiber patch cords and optical components before the 2$\times$1 switch and after the DWDM need to be compensated once in an initialization/characterization process. This is done by inserting a QWP, HWP, and phase retarder in each of these sections and adjusting them until non-orthogonal input test states (V) and (D) propagate through unperturbed (demonstrating no birefringence in these short sections). We use FPCs to serve as the compensating QWPs and HWPs and insert either LCRs or birefringent plates for the required phase retardance. Fibers and fiber components in these sections are then carefully tacked down so that the birefringence effectively remains constant through the experiment. Because total fiber lengths are short in these sections (<\,50\,m), the birefringence changes of environmental fluctuations (temperature and vibration) are negligible. The only non-negligible, changing, birefringence therefore comes from environmental fluctuations of the fiber link under test and/or our intentional sudden polarization changes made by the commercial polarization controller to emulate changing fiber routes on a multi-node network. In this way, once the system is initially stabilized, different fiber links and optical components can be interchanged between the 2$\times$1 switch and the DWDM. The compensator then updates to ensure that input polarization states are faithfully transmitted to the output, provided the system operates within its tolerances.     

\subsection{Calibration and Intrinsic Stabilization Accuracy}\label{subsection:baseline}
 
\begin{figure}[h!]
\centering\includegraphics[width=\linewidth]{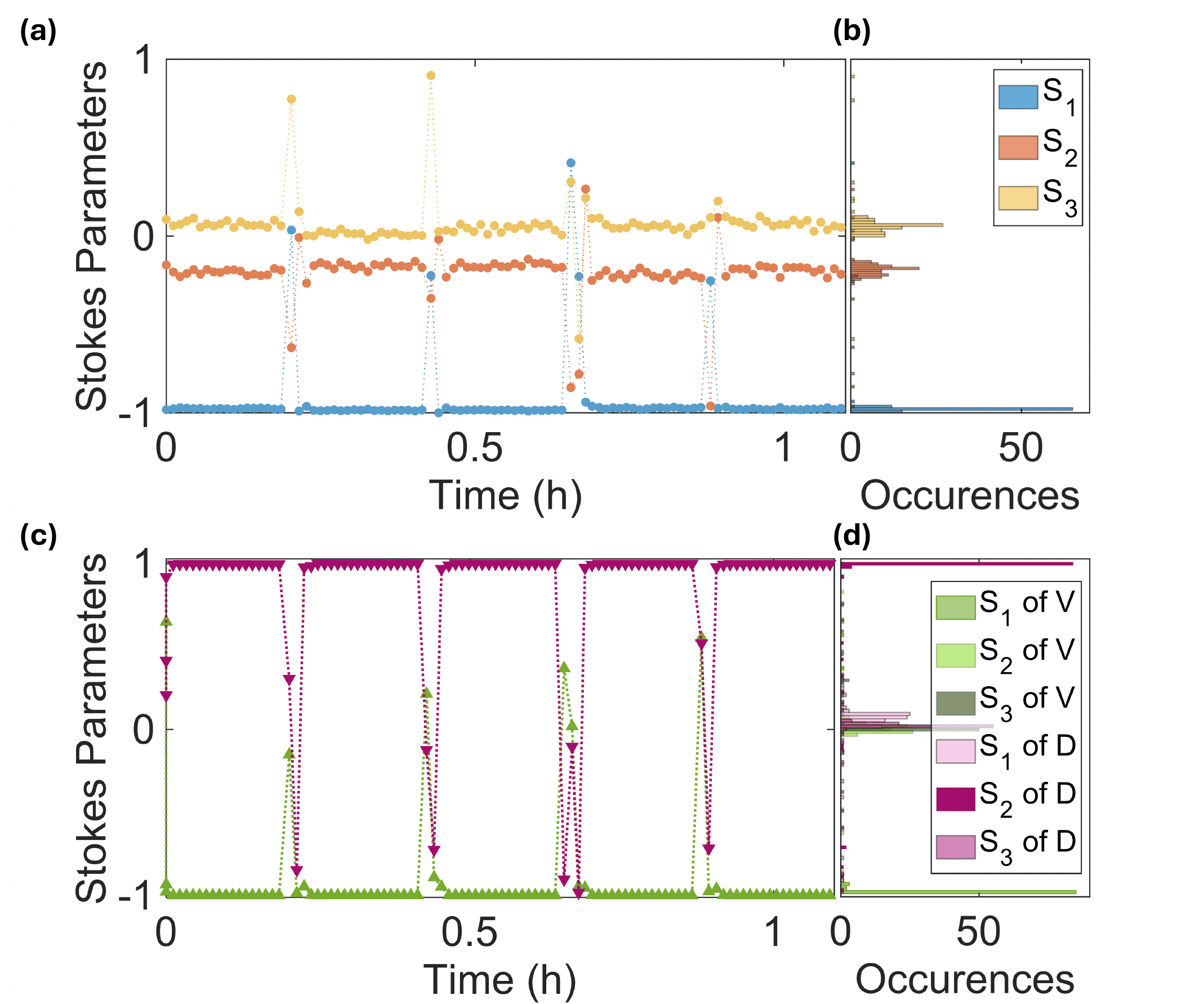}
\caption{\remove{(a) Experimental setup illustrating the use of wavelength-detuned classical headers for polarization monitoring and active compensation. The commercial polarization controller is employed to mimic rerouting in the network.} 
(\remove{b}\add{a}) Measured Stokes parameters $S_1$, $S_2$, and $S_3$ of the polarized single photons over time, showing stable polarization with occasional changes due to the commercial polarization controller. The Stokes parameters are calculated from single photon counts measured in the H, V, D, and R polarization bases.
(\remove{c}\add{b}) Histogram of the Stokes parameters in (b), demonstrating tight distributions and long-term polarization compensation after intentional changes.
(\remove{d}\add{c}) Time evolution of the Stokes parameters for vertical (V) and diagonal (D) reference states, confirming reliable recovery of nonorthogonal polarization states.
(\remove{e}\add{d}) Corresponding histograms of the Stokes parameters for the V and D states, indicating robust stabilization performance and minimal drift over extended operation.
}
\label{fig:result_no_fiber_single_pol}
\end{figure}

Before showing birefringence compensation on deployed fiber, we calibrate the system back-to-back without a fiber link under test. In order to emulate a realistic multi-node network with various different fiber routes a packet might take (and therefore multiple sudden changes in birefringence over quantum datagram transmission), we place a programmable polarization controller between the switch and the fiber, and periodically switch the input polarization into the fiber to a random state\remove{, Fig.~\ref{fig:result_no_fiber_single_pol}(a)}. In this way, every update of the polarization controller gives a new effective fiber transfer matrix, simulating a new fiber route. We automatically pick a rotation angle between 0 and $\pi$ from the uniformly distributed interval of [1,999] for each pedal of the four-pedal polarization controller. The integer numbers 1 and 999 correspond to 0 and $\pi$, respectively. We repeat this process four times over an hour so that our protocol compensates for a four effective different fiber routes per hour. 

\remove{The quantum payload for these experiments is a single polarized photon created via cascaded $\chi^{(2)}:\chi^{(2)}$ nonlinear processes at 1309\,nm.}\add{The quantum payload in these experiments is 1318\,$\mu$s long and consists of the signal photons of correlated photon pairs generated in a single periodically poled lithium niobate waveguide (PPLN) through a cascaded $(\chi^{(2)}:\chi^{(2)})$ process. In this source, 1320\,nm pump pulses (100\,ps pulse width, 500\,MHz repetition rate) undergo second-harmonic generation (SHG), and the generated second-harmonic field subsequently drives spontaneous parametric down-conversion (SPDC) within the same waveguide to produce correlated signal and idler photon pairs. Because the quantum payload duration is much longer than the pump pulse period (2\,ns), each quantum payload can potentially contain many signal photons over the payload interval, depending on the photon pair generation rate in our system.} \add{We use superconducting nanowire single-photon detectors (SNSPD) for quantum payload measurements} We ignore the idler photons for this test as the polarization compensation is only employed in the signal arm. By projecting the signal photons onto four different polarization states, horizontal (H), vertical (V), diagonal (D), and right-circular (R), we calculate the Stokes parameters of the signal photons as [$-$0.98,\;$-$0.16,\;0.09], corresponding to normalized Stokes parameters [$S_1$,\;$S_2$,\;$S_3$].  Figure~\ref{fig:result_no_fiber_single_pol}(\add{a}) shows the evolution of the Stokes parameters over 1 hour. They are stable with an average angular distance of \ang{2.8} on the Poincare sphere as shown in Fig.~\ref{fig:result_no_fiber_single_pol}(\add{b}) unless we intentionally introduce a new polarization transfer matrix by changing the state of the polarization controller. Figure~\ref{fig:result_no_fiber_single_pol}(\add{c}) and Fig.~\ref{fig:result_no_fiber_single_pol}(\add{d}) demonstrates the time evolution of the Stokes parameters of the two non-orthogonal polarizations states of the headers that we use for active compensation. Their ideal states are $-$1(1) for $S_1$($S_2$) of V(D) polarizations. We set the thresholds to enter a polarization compensation cycle if $S_1$ of rotated state $S_V^\prime<$\,$-$0.999 and $S_2$ of rotated state $S_D^\prime<$\,0.995. According to these thresholds, the angular distance of the rotated vertical state $S_V^\prime$ from the ideal vertical state on the Poincare sphere (arccos($S_{\mathrm{i}}\cdot S(t))$)  is predicted to be no more than \ang{2.5} and that for the rotated diagonal state $S_D^\prime$ from its ideal diagonal state to be no more than \ang{5.7}. Back-to-back measurements of Stokes parameters for \remove{single photons}\add{headers}, Fig.~\ref{fig:result_no_fiber_single_pol}(\add{c}) and (\add{d}) correspond to angular distances from their ideal values of less than \ang{1.5} for (V) and less than \ang{1.7} for (D) unless we purposely disrupt them with the polarization controller. These small angular distances indicate that not only our fiber patch-cords in the lab and interfacing optical components are stabilized to the environment, but that classical reference headers in our setup indeed accurately track the birefringence/polarization deviation of single photons (i.e. header stabilization shown in Fig.~\ref{fig:result_no_fiber_single_pol}(\add{c}) and (\add{d}) correspond to stabilized photons shown in Fig.~\ref{fig:result_no_fiber_single_pol}(\add{a}) and (\add{b}). We also note that the error in quantum measurements is limited by the square root of the number of photons detected based on the single photon detection statistics. In these measurements, the maximum photon counts were detected in vertical polarization with 50 kilocounts per second (kcps). Lower photon count rates will therefore show less reciprocity between stabilized photons and stabilized headers that are not due to tracking of our system but rather from counting error. 

\begin{figure}[h]
\centering\includegraphics[width=\linewidth]{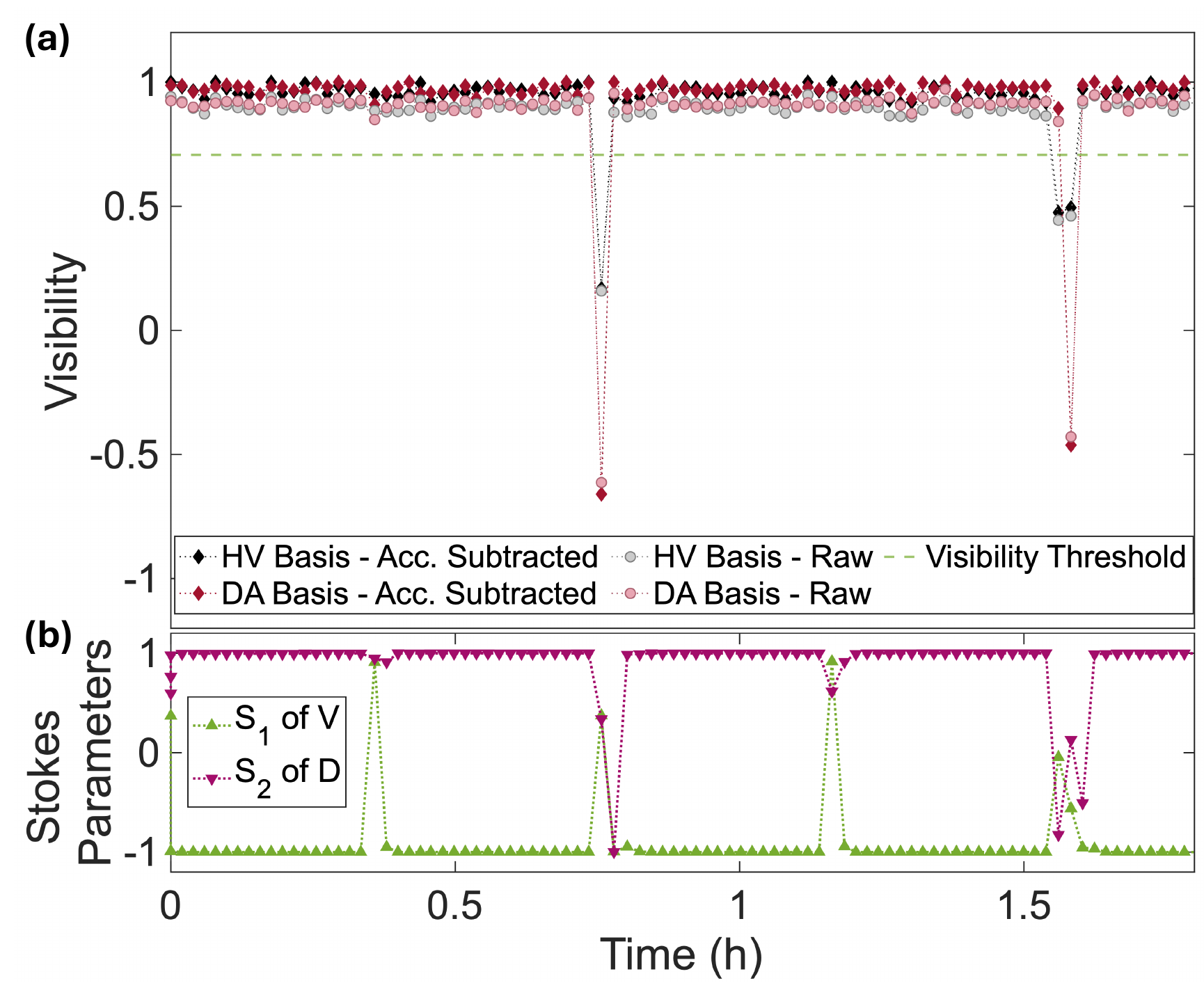}
\caption{(a) Measured \remove{two-photon interference}\add{TPI} visibility in the horizontal–vertical (HV) and diagonal–antidiagonal (DA) bases over 2\,hours (circles), and after accidental subtraction (diamonds). High visibility is maintained unless there is a change we intentionally introduced by the commercial polarization controller. Notably, visibility degradation is observed during the second and fourth recovery rounds but not during the first and third, as the vertical and diagonal reference states remain significantly displaced from their target values after the first compensation cycle. (b) Time evolution of the relevant Stokes parameters for the vertical (V) and diagonal (D) reference states, demonstrating that they return their ideal states within the thresholds we set. 
}
\label{fig:result_no_fiber_vis}
\end{figure}

Next, we characterize the system using entangled photon pairs. \remove{In these experiments, we prepare photons in the Bell state, $\ket{\Phi^{+}}=\frac{1}{\sqrt{2}}\left(\ket{\mathrm{HH}}+\ket{\mathrm{VV}}\right)$, and}
\add{In these experiments, we configure the correlated photon pair source in a Sagnac loop to prepare polarization-entangled photon pairs in the Bell state, $\ket{\Phi^{+}}=\frac{1}{\sqrt{2}}\left(\ket{\mathrm{HH}}+\ket{\mathrm{VV}}\right)$.} We measure the visibility bounds in the HV basis using \add{$V_{\mathrm{HV}} = (C_{\mathrm{HH}}-C_{\mathrm{HV}})/(C_{\mathrm{HH}}+C_{\mathrm{HV}})$} over 2\,hours, Fig.~\ref{fig:result_no_fiber_vis}. We repeated the measurements for the DA basis where we use D and A instead of H and V in the equation above. During this experiment, we also intentionally introduced abrupt birefringence changes with the polarization controller (to emulate changes in fiber path from network rerouting) and observed the ability of the system to compensate. 

The average visibilities over the first 20 measurements are 91.0$\pm$2.1\% in the HV basis and 91.5$\pm$1.2\% in the DA basis, prior to any rerouting and recovery trials (i.e. the system is left to drift for 20 measurements after an initial system birefringence stabilization). Over the entire experiment, the visibilities are measured as 90.3$\pm$2.0\% for the HV basis and 91.6$\pm$1.6\% for the DA basis, unless we intentionally force the system outside the predefined thresholds using the commercial polarization controller. These values correspond to the raw visibilities with a coincidence-to-accidentals ratio \add{($CAR = (CC-ACC)/ACC$)} of 25$\pm$1. Based on the CAR measurements, the visibility of the source is limited to $\sim$92.6\% by accidental coincidences due to multiple pair generation \cite{takesue_effects_2010,takeoka_full_2015,brewster_quantum_2021}. Accidental counts are estimated by averaging over 200 time-shifted integration windows in the coincidence count histogram, each separated by 2\,ns from the true coincidence peak and integrated over a 1\,ns window. Negative coincidence values arising from accidental subtraction due to statistical fluctuations are set to zero before calculating the visibility, ensuring physically meaningful results. In Fig.~\ref{fig:result_no_fiber_vis}, after subtracting accidentals, the initial visibilities improve to 97.2$\pm$2.1\% and 97.9$\pm$1.2\%, and remain at 96.4$\pm$2.1\% and 97.6$\pm$1.6\% in the HV and DA bases, respectively. The negligible changes in the visibilities indicate that information-carrying classical header bits can be used for polarization tracking and active compensation. In the next section, we test single- and entangled-photon performance over a deployed fiber link of 48\,km. Here, due to the much longer interaction length with the fiber compared to the back-to-back calibration, environmental birefringence fluctuations are not negligible and must be compensated. Additionally, the ability of headers to accurately probe the fiber transfer matrix degrades with fiber length and channel wavelength separation due to PMD.

\subsection{Deployed Fiber Link Under Test}\label{subsection:link_characterization}

\begin{figure}[h]
\centering\includegraphics[width=\linewidth]{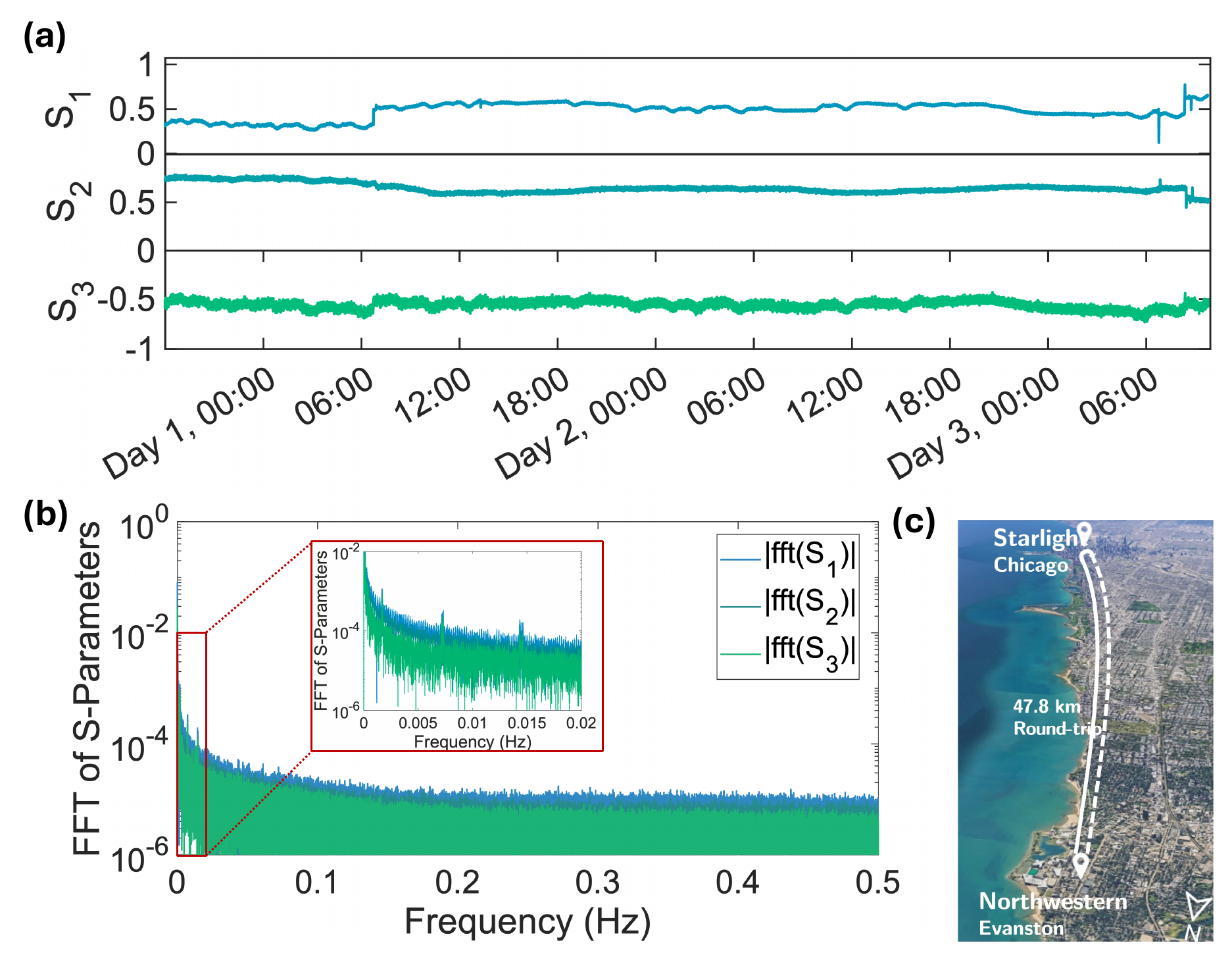}
\caption{Polarization stability of the deployed Chicago fiber loop.
(a) Time evolution of the normalized Stokes parameters $S_1$, $S_2$, and $S_3$
measured over approximately 60\,hours for a 47.8\,km round-trip deployed fiber
link between Northwestern University (Evanston) and the Starlight node in
downtown Chicago. Slow drifts are observed, reflecting
environmentally induced slow polarization changes in the field-deployed fiber.
(b) Magnitude of the Fourier transform of the Stokes parameters,
$|\mathrm{FFT}(S_1)|$, $|\mathrm{FFT}(S_2)|$, and $|\mathrm{FFT}(S_3)|$,
showing that the slow polarization fluctuations occurring at low frequencies
below 0.5\,Hz. These experiments cannot resolve higher-frequency components due to the sampling rate of our device. The inset highlights the low-frequency spectral content.
(c) Map of the deployed fiber loop, indicating the 47.8\,km round-trip path between Evanston and Chicago. Map data: Google Earth (\textcopyright~Google).}
\label{fig:fiber_link}
\end{figure}

The fiber link under test lies between the Northwestern University Evanston campus and StarLight in downtown Chicago. It is looped back at StarLight by an 8$\times$8 switch to create a total of 47.8\,km underground fiber link. The performance of our method in this metropolitan scale system depends on the attenuation and total polarization mode dispersion (PMD). This fiber link has a total attenuation of 19\,dB at 1310\,nm due to the insertion loss of the single mode fibers, the switch at the StarLight location, and splicing and connector losses. The attenuation has not been a factor in our measurements thus far partly because the sensitivity of our polarization measurement is enhanced by the PDFA. Sensitive detectors such as avalanche photodiodes or SFPs with higher power budgets can also be used to limit the impact of attenuation. In Fig.~\ref{fig:fiber_link}(a), the normalized Stokes parameters of the deployed fiber link are shown over two days. 
These measurements are taken by \add{launching a CW laser at 1300\,nm through a fiber polarizer and the deployed 47.8\,km link. The evolution of the polarization state of the transmitted classical light is monitored using a Thorlabs PAX polarimeter at a sampling rate of 100\,samples/s}. The corresponding Fourier spectra in Fig.~\ref{fig:fiber_link}(b) reveal that the polarization fluctuations are dominated by low-frequency components below $\sim$0.02\,Hz, with negligible amplitudes at higher frequencies. This indicates that polarization evolution in the deployed loop occurs on timescales of tens of seconds to minutes. Such low-frequency behavior is favorable for practical quantum networking systems since it enables reliable polarization tracking and compensation. These measurements establish the baseline polarization dynamics of the deployed fiber and motivate the use of header-based polarization monitoring and stabilization for robust long-haul quantum network operation.

\section{Experimental Results on Deployed Fiber}\label{section:results}
\subsection{Channel Stabilization Data for Single Photon Qubits}\label{subsection:loop_single_photon}

\begin{figure}[h!]
\centering\includegraphics[width=\linewidth]{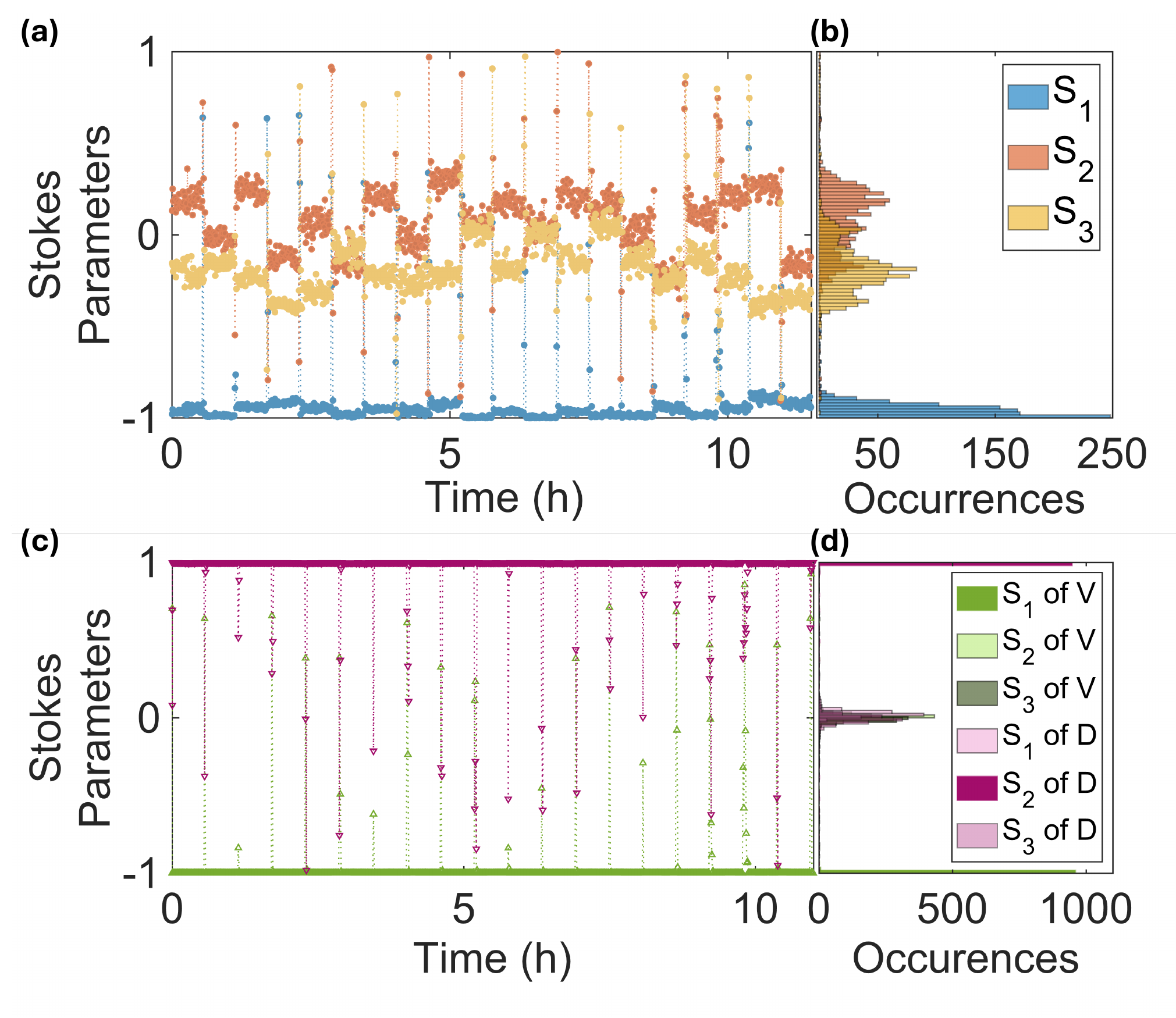}
\caption{
(a) Time evolution of the measured Stokes parameters \(S_1\), \(S_2\), and \(S_3\) of the single photon qubits, prepared in state [$-$0.98,\;0.13,\;$-$0.16].
(b) Histogram of the Stokes parameters in (a), indicating a spread compared to the baseline measurements for sudden and random changes in the polarization transfer matrix.
(c) Time evolution of the Stokes parameters for the vertical (V) and diagonal (D) reference states, demonstrating reliable recovery of nonorthogonal polarization states throughout extended operation regardless of the loss due to the fiber link.
(d) Corresponding histograms of the Stokes parameters for the V and D states, confirming robust long-term stabilization performance.
}
\label{fig:sudden_change_single_pol_chi}
\end{figure}

By using these results as our baseline, we repeated these tests on the deployed fiber link described in Sec.~\ref{section:setup}\ref{subsection:link_characterization}. Figure~\ref{fig:sudden_change_single_pol_chi}(a) shows the evolution of Stokes parameters of the quantum payload projected on a polarization state with Stokes parameters of [$-$0.98,\;0.13,\;$-$0.16]. For compensation, we expect Stokes parameters not to deviate from the initial values after the "routing and recovery" cycles. With active compensation, the quantum signal must return to the target photon count rate after intentional polarization scrambling using a fiber polarization controller before the fiber spool.
Both Fig.~\ref{fig:sudden_change_single_pol_chi}(a) and (b) show that there is a deviation from the target value even though the header polarization states return to the ideal values after each cycle as shown in Fig.~\ref{fig:sudden_change_single_pol_chi}(c) and (d). We calculate the average angular distance from the target state [$-$0.98,\;0.13,\;$-$0.16] as \ang{10.8} with a standard deviation of \ang{5.8}. This is higher than the angular distance from the ideal state we observed without the fiber link. We should also note that in this experiment the average angular distance is \ang{5.1} with a standard deviation of \ang{3} even without the "rerouting and recovery" trials (i.e. the deviation within each "flat" section where classical header probes do not require the system to go into compensation, Fig.\ref{fig:sudden_change_single_pol_chi}(c) and (d). This degradation of the polarization compensation performance between compensation cycles at each round can be explained by the polarization mode dispersion (PMD) in fibers which rotates the headers and quantum payload slightly differently due to the wavelength detunings between them (see Sec.~\ref{section:theory}). Related, the deviation of the levels themselves corresponds to the random states prepared by the polarization controller being more or less aligned with the PSP axes of the transfer matrix of the fiber. For example, when the output state of the polarization controller is aligned closer to one of the PSPs, the transfer matrix calculated from classical headers more faithfully reproduces that of the quantum payload and the levels will be closer to the ideal state [$-$0.98,\;0.13,\;$-$0.16]. The largest deviations correspond to "unlucky" input polarizations where the random value corresponds to input states that are close to being orthogonal to the PSPs of the fiber. 

\subsection{Channel Stabilization Data for Entangled Photon Qubits}\label{subsection:chi_loop_visibility}

\begin{figure}[h!]
\centering\includegraphics[width=1\linewidth]{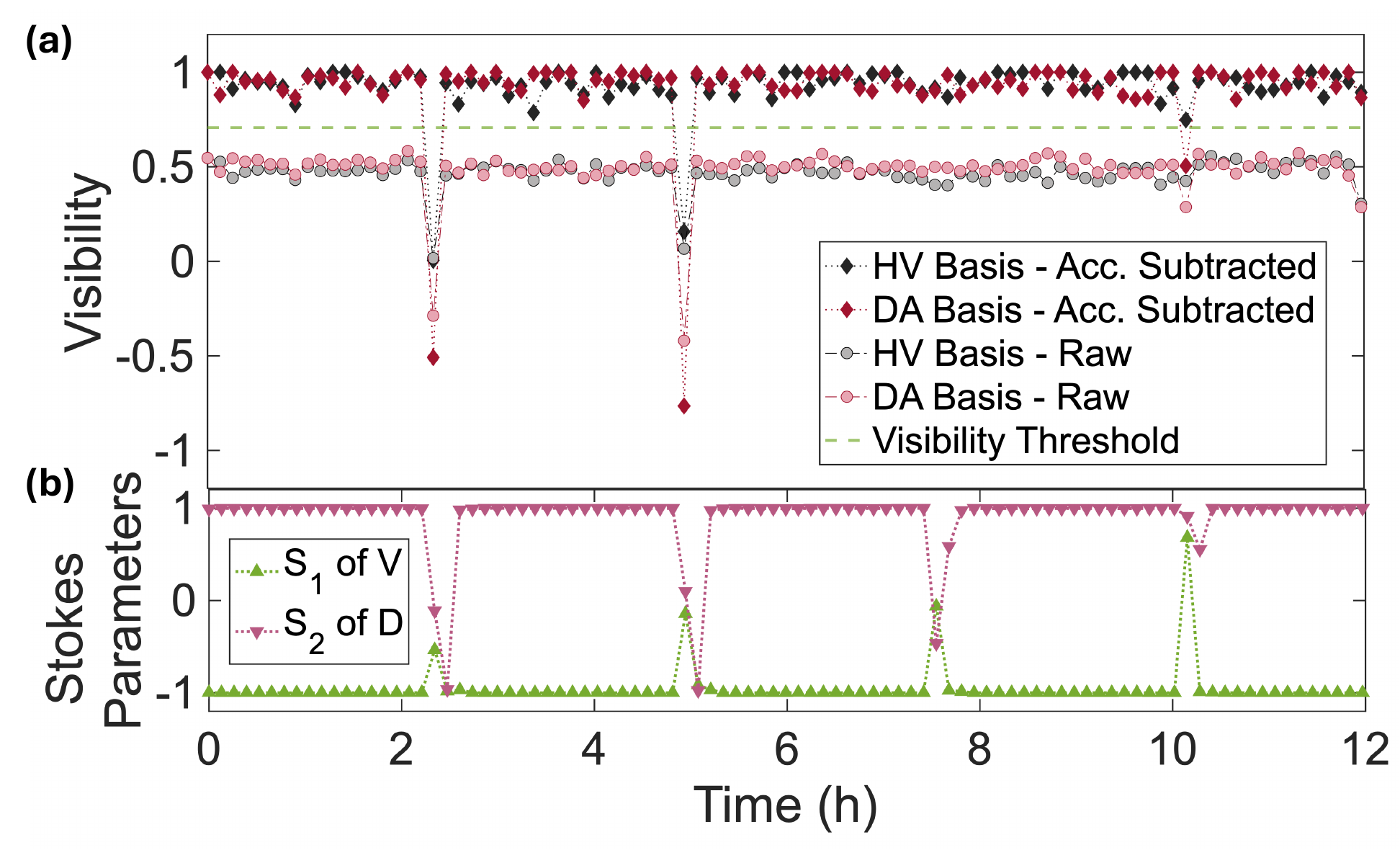}
\caption{
(a) Measured \remove{two-photon interference}\add{TPI} visibility bounds in the horizontal--vertical (HV) and diagonal--antidiagonal (DA) bases over a 12 hours. The visibility is maintained above 70.7\% unless we intentionally introduce a change in birefringence. (b) Time evolution of the relevant Stokes parameters for the vertical (V) and diagonal (D) reference states, showing that they return to their ideal states within the thresholds we set. }
\label{fig:result_Chicago_loop_vis}
\end{figure}

Figure~\ref{fig:result_Chicago_loop_vis} shows visibility tests on the deployed fiber link for entangled photons prepared in the Bell state, $\ket{\Phi^{+}}=\frac{1}{\sqrt{2}}\left(\ket{\mathrm{HH}}+\ket{\mathrm{VV}}\right)$. The average raw visibilities over the first 20 measurements are 48.7$\pm$2.9\% in the HV basis and 52.2$\pm$3.2\% in the DA basis, prior to any rerouting and recovery trials. Over 12\,hours, the visibilities are measured as 47.5$\pm$3.4\% for the HV basis and 51$\pm$3.2\% for the DA basis. These raw visibilities  are limited to 50.5\% by accidental coincidences. Accidental counts are estimated with the same procedure described in Sec.~\ref{section:setup}\ref{subsection:baseline}. In Fig.~\ref{fig:result_Chicago_loop_vis}, after subtracting accidentals, the initial visibilities improve to 95.6$\pm$4.5\% and 95.1$\pm$4.4\%, and remain at 94.2$\pm$5\% and 95.4$\pm$4.6\% in the HV and DA bases, respectively. 

\subsection{Continuous 44-Hour Stabilization Performance}\label{subsection:loop_long_term}

\begin{figure}[h!]
\centering\includegraphics[width=0.85\linewidth]{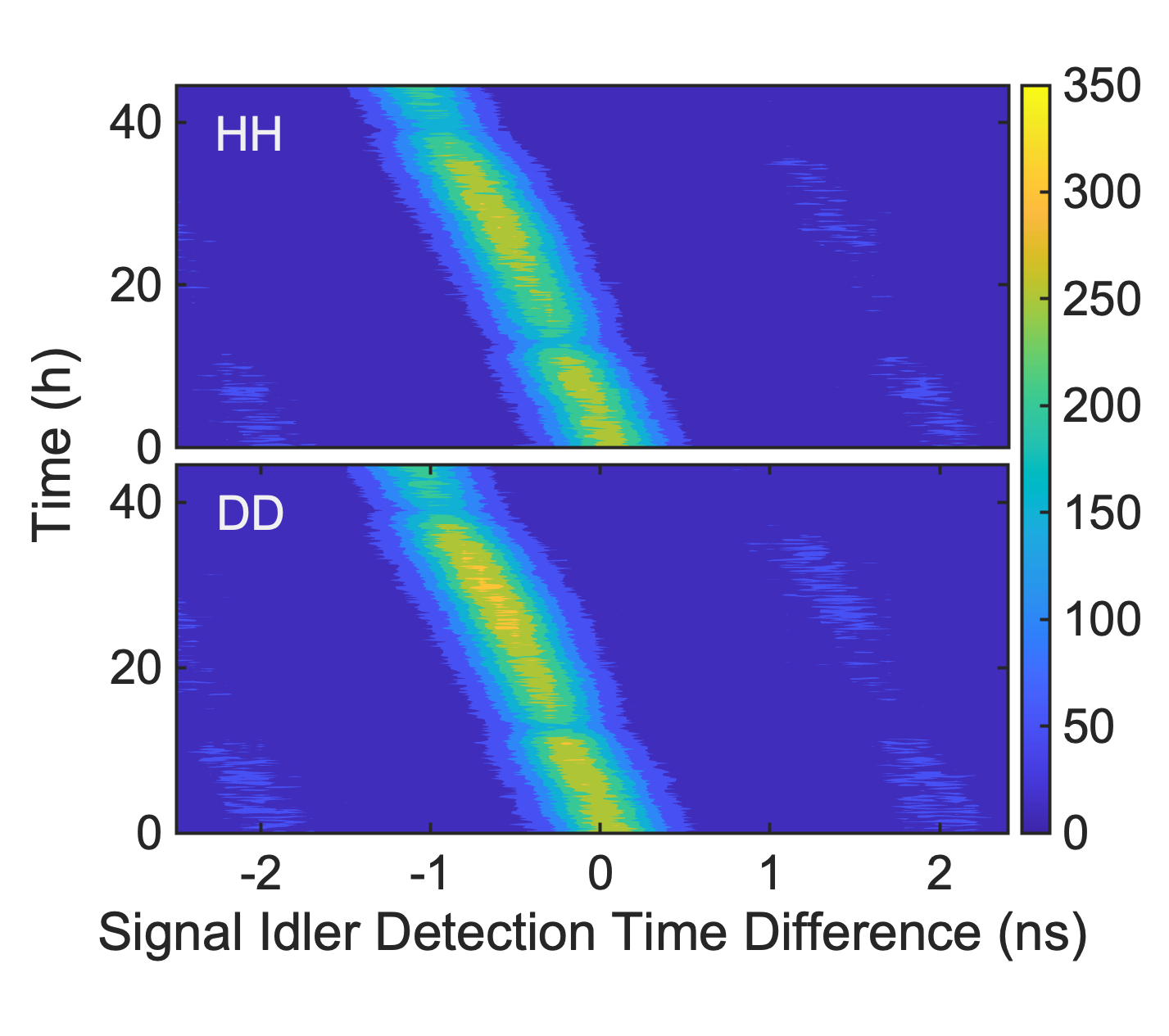}
\caption{Coincidence counts measured in the $\ket{\mathrm{HH}}$ and $\ket{\mathrm{DD}}$ polarization bases for the signal and idler photons over a 44-hour continuous acquisition. The histograms have 100\,ps bin width. Active polarization compensation was disabled at hour~32 after which the coincidence counts decrease. A gradual shift to earlier delays of the coincidence peak is observed over time with and without the polarization compensation, indicating an environmental change impacting the arrival time such as a variation in temperature.}
\label{fig:long_tem_3D}
\end{figure}

We additionally measure the performance of birefringence channel compensation under slow, long-term, continuous drifts. The data in previous sections indicated that classical headers rarely trigger a compensation cycle over time scales of $\sim$1\,hr for the deployed fiber we use. However, in a realistic network setting, one will also need to compensate larger, slower variations due to environmental fluctuations such as drastic temperature variations from large weather changes. For these tests, we prepare entangled photons in the state $\ket{\Phi^{+}}=\frac{1}{\sqrt{2}}\left(\ket{\mathrm{HH}}+\ket{\mathrm{VV}}\right)$ and measure visibility over 44 hours, Fig.~\ref{fig:long_tem_3D}. The active polarization compensation was disabled after 32 hours to look at effects from natural drift of the system over the last 12 hours. 

\begin{figure}[h!]
\centering\includegraphics[width=\linewidth]{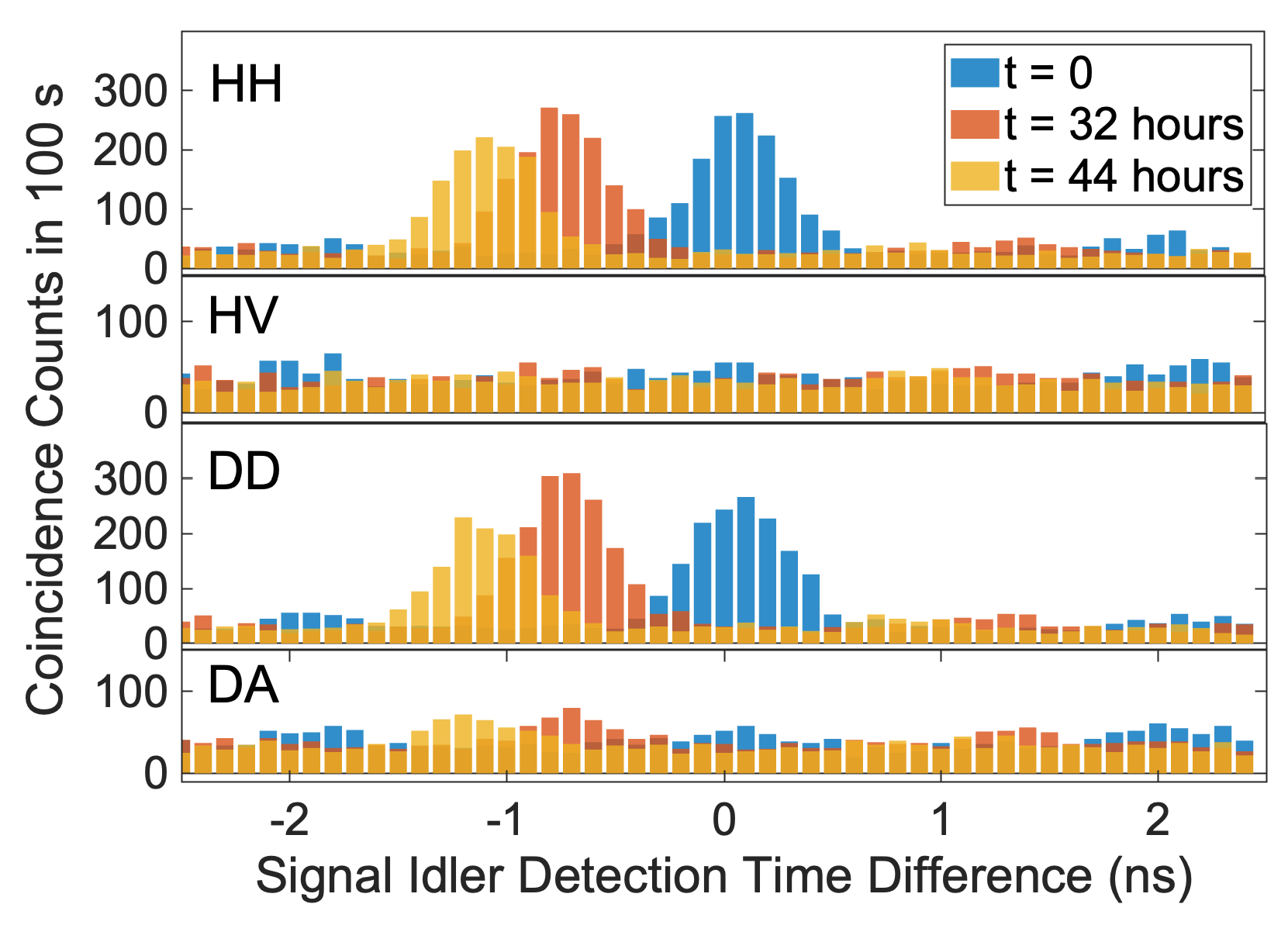}
\caption{Signal--idler coincidence counts histograms measured over a 100\,s integration window in different polarization bases during a 44-hour continuous experiment. The histogram bins have a width of 100\,ps. Panels show measurements in the $\ket{\mathrm{HH}}$, $\ket{\mathrm{DD}}$, $\ket{\mathrm{HV}}$, and $\ket{\mathrm{DA}}$ bases. Coincidence histograms are shown at the start of the experiment at $t=0$, at $t=32$~hours, and at $t=44$~hours. Active polarization compensation was disabled at $t=32$~hours, after which the number of coincidence counts decrease. No distinct coincidence peak is observed in the orthogonal bases $\ket{\mathrm{HV}}$ whereas the small peak for $\ket{\mathrm{DA}}$ shifts to earlier arrival times similar to (a) and (b).}
\label{fig:long_term_measurement}
\end{figure}

Figure~\ref{fig:long_term_measurement} shows three moments from the long term measurement in Fig.~\ref{fig:long_tem_3D}: the initial measurements at t = 0, the last measurement before the polarization compensation is deactivated at t = 32 hours and the last measurement of this set at t = 44 hours. 

\begin{figure}[h!]
\centering\includegraphics[width=\linewidth]{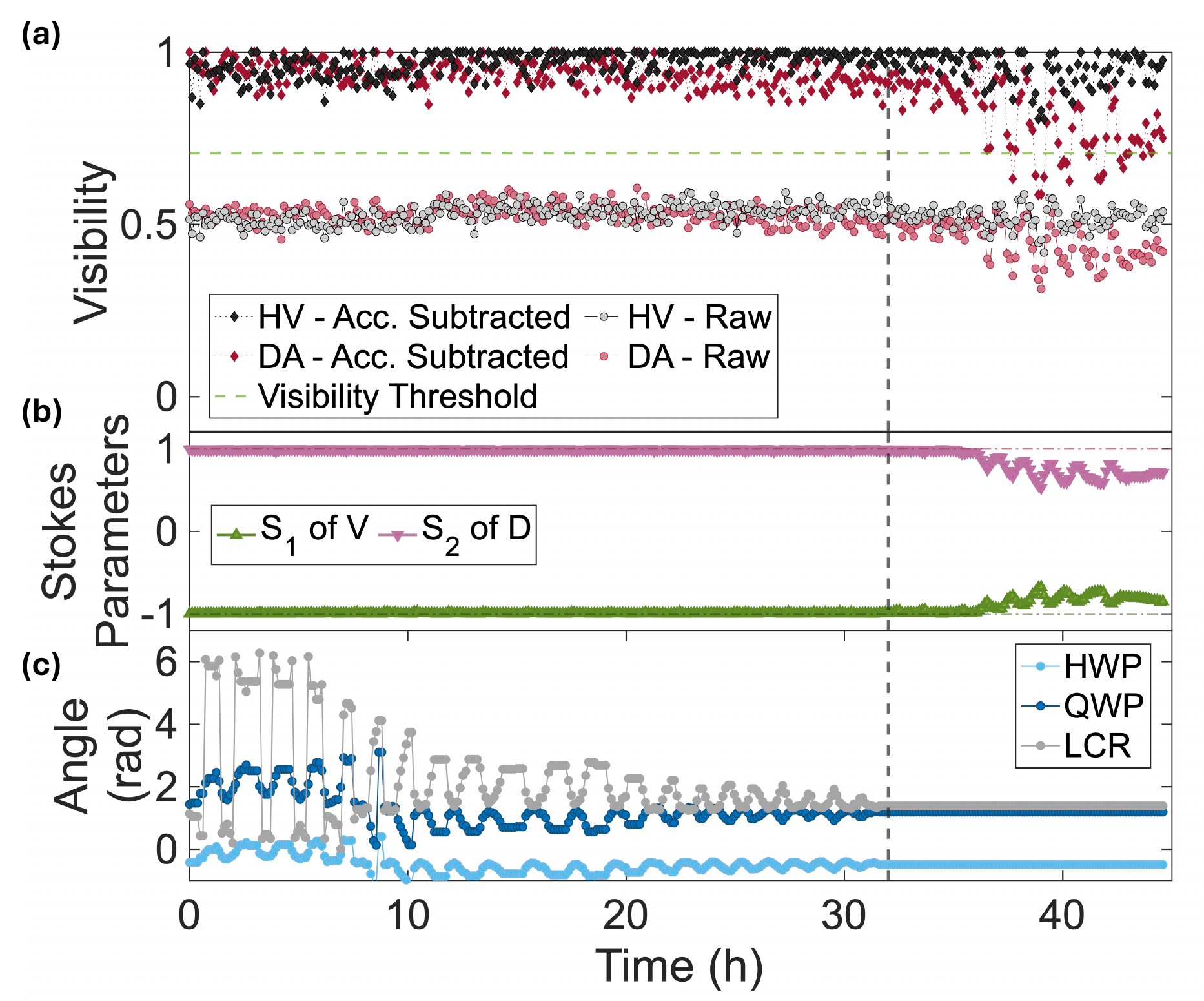}
\caption{Long-term stability of polarization correlations over a 44-hour continuous experiment. (a) Raw \remove{two-photon}\add{TPI} visibility (circles) measured in the HV and DA bases as a function of time. Corresponding visibilities after accidental coincidence subtraction (diamonds). The green dashed line at 0.71 indicates the threshold visibility for quantum. The vertical dashed line at $t=32$~hours indicates the time at which active polarization compensation was disabled, after which increased visibility fluctuations are observed, particularly in the DA basis. (b) Measured Stokes parameters $S_1$ of the V state and $S_2$ of the D state, showing stable polarization tracking prior to disabling compensation and increased drift thereafter. (c) Evolution of the angles of the half-wave plate (HWP), quarter-wave plate (QWP), and the phase provided by the liquid crystal retarder (LCR) during the experiment, illustrating the stabilization dynamics and the static behavior after compensation is turned off.}

\label{fig:visibility}
\end{figure}

In addition to an observed 1\,ns drift to earlier arrivals of the coincidence counts peak location in 44 hours, the number of coincidence counts also decreased. However, we also observe a decrease in coincidence counts around hour 13 (with compensation) and hour 39 (without compenstaion). This lack of a discernible pattern indicates that coincidence counts alone cannot predict the impact of polarization drift. 
Figure~\ref{fig:visibility}(a) shows that during compensation, the raw visibilities on the HV and DA bases are both 53.0$\pm$2.7\%. They decrease to 52.1$\pm$3.0\% and 45.3$\pm$5.7\%, respectively, when compensation is disabled. \remove{The visibilities after the subtraction  of accidental counts are obtained by averaging the coincidence counts over 200 off-peak time windows, each shifted by 2\,ns from the main coincidence peak and integrated over a 1\,ns duration similar to the baseline measurement in Sec.~\ref{section:setup}\ref{subsection:baseline}.}\add{The visibilities after accidental-count subtraction are obtained using the procedure described in Sec.~\ref{section:setup}\ref{subsection:baseline}.} The average visibility in the HV basis decreases from 96.8$\pm$3.4\% to 95.3$\pm$4.7\% when the polarization compensation is disabled. Similarly, in the DA basis, the accidental-subtracted visibility drops from 93.8$\pm$ 3.7\% to 81.0$\pm$ 10.2\%. 

Though the decrease in visibility from long-term environmental drift is not large compared to the sudden changes we use to emulate packet rerouting in Sec.~\ref{section:results}\ref{subsection:chi_loop_visibility}, it is clear that the compensation module is regularly undergoing cycles to continually negate the changing birefringence of the channel up until hour 32. To illustrate the correlation between classical header polarization and \remove{two-photon interference}\add{TPI} visibilities, we show the corresponding Stokes parameters of classical headers in Fig.~\ref{fig:visibility}(b). After the compensation is disabled, Stokes parameters begin drifting from their ideal values and visibilities simultaneously degrade. The angles of the waveplates and the voltage of the LCR in compensation modules remain at their final values after hour 32 as shown in Fig.~\ref{fig:visibility}(c). Clearly, classical headers accurately probe the degradation in the quantum channel during long-term environmental drift just as they did for large sudden changes. 

\subsection{Compensation Time Statistics and Threshold Effects}\label{subsection:comp_time}
\begin{figure}[h!]
\centering\includegraphics[width=0.8\linewidth]{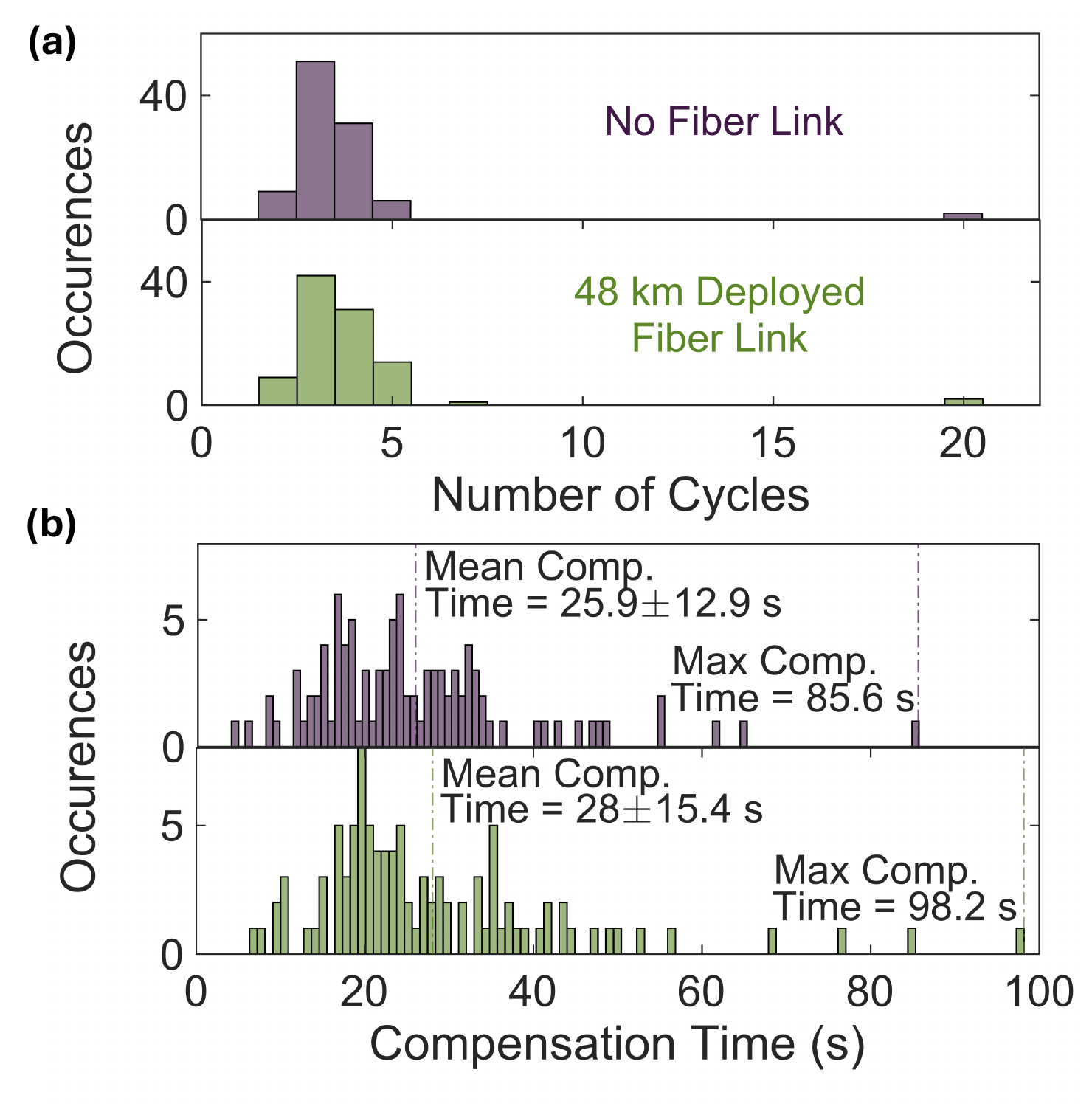}
\caption{Statistical characterization of polarization compensation performance measured over 100 trials. (a) Histogram of the number of compensation cycles required to reach convergence in the absence of a fiber link (purple) and with a 48\,km deployed fiber link (green). (b) Distribution of the corresponding compensation times for the absence of a fiber link (purple) and with a 48\,km deployed fiber link (green).}
\label{fig:comp_speed}
\end{figure}

We evaluate whether the length of the fiber link affects the speed of the active polarization control system through loss or reduced signal-to-noise ratio (SNR). For this purpose, we compare the compensation duration measured in the baseline configuration with that measured using the deployed 48 km fiber link in the Chicago area (see Sec.~\ref{section:setup}\ref{subsection:link_characterization}). In this study, we employ the commercial polarization controller and randomly change the polarization state into the fiber link 100 times. As shown in Fig.~\ref{fig:comp_speed}(a), the number of cycles required to compensate for the polarization change is 2 to 5 cycles. In other words, the classical header reference states are typically returned from their rotated values to their target vertical and diagonal states within 2–5 cycles. The average compensation time remains nearly identical in the cases with and without the deployed link, 25.9 and 28 s respectively. This indicates that the compensation dynamics are primarily determined by the time required for the motorized stages to apply the rotation angles, rather than the transmission distance. This behavior verifies that the compensation module is indeed independent from the quantum measurements and the number of photons detected as expected.

As more fiber and/or other loss mechanisms are added to the system, the SNR of the classical headers measured at the inline polarimeter will decrease.
At some point, the SNR becomes too low to provide accurate information about the fiber transfer function which is used to calculate the compensator rotations. However, for the 48\,km link, the system performance for timing and stability of classical header control is nearly as good as the back-to-back case with no fiber.  

\begin{figure}[h!]
\centering\includegraphics[width=\linewidth]{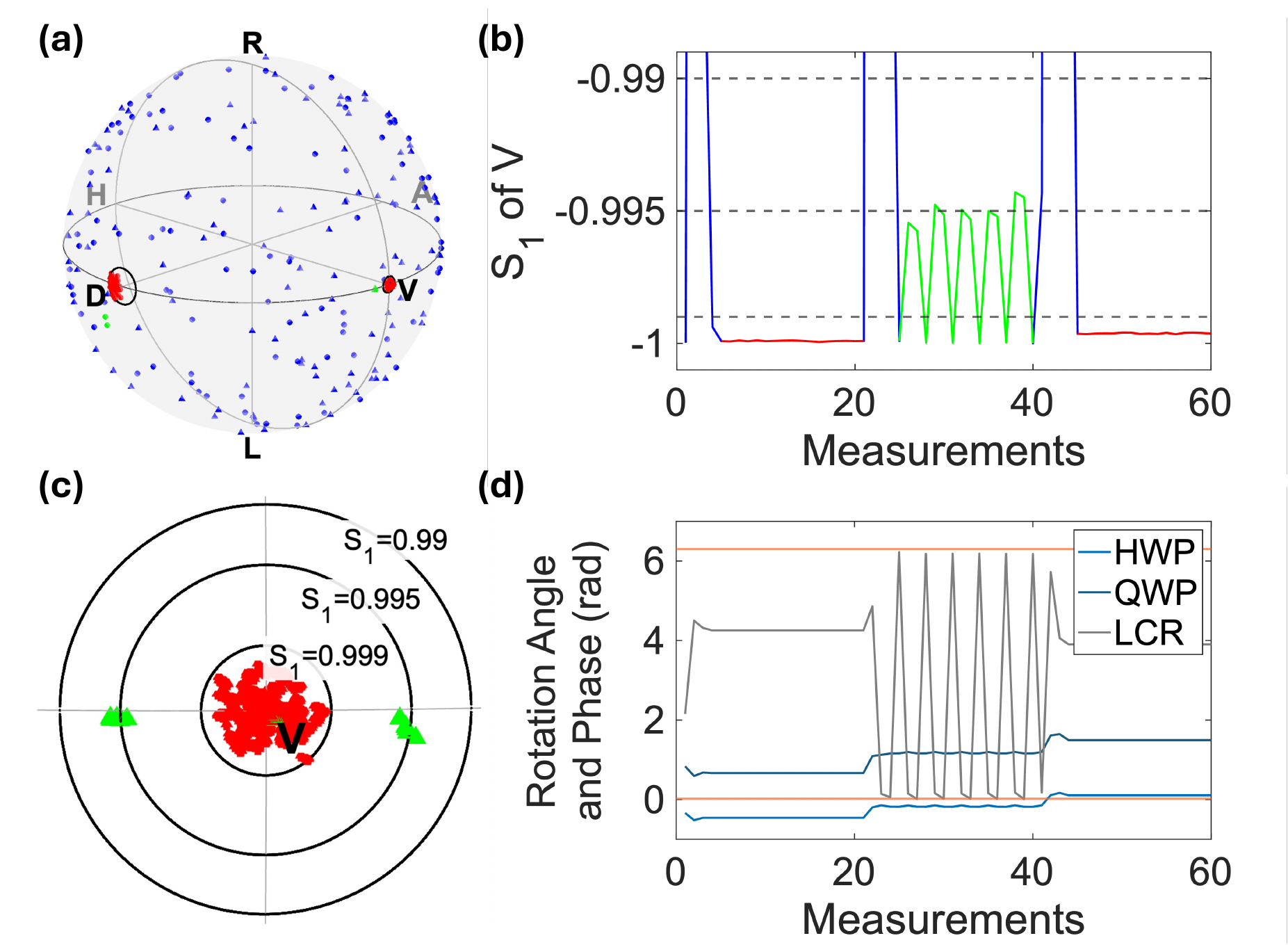}
\caption{Representative example of a non-convergent polarization compensation trial. (a) \remove{Measured polarization states projected on the Poincar\'e sphere during the compensation process. The target vertical (V) and diagonal (D) states are indicated with circles around them. Blue dots corresponds to the cycles where compensation module actively compensating. The red dots correspond to satisfied conditions and compensation is not necessary.}\add{Measured polarization states projected on the Poincar\'e sphere before and after the compensation process. The target vertical (V) and diagonal (D) states are indicated with circles around them. Blue triangles and dots correspond to the cycles before the compensation begins. The red triangles and dots correspond to satisfied conditions, and green triangles and dots represent 2 out of 100 cases where the conditions are not satisfied.}  
(b) Evolution of the Stokes parameter $S_1$ of the V state over successive measurements in a non-convergent compensation cycle. The dashed horizontal lines indicate possible tolerance bounds where the lowest line, $-$0.999, corresponds to the tolerance used in our experiments. 
(c) Projection of the V state on the Poincar\'{e} sphere with concentric contours corresponding to $S_1=0.99$, $0.995$, and $0.999$, illustrating that the compensated states remain outside the target tolerance region but close. 
(d) Corresponding evolution of the half-wave plate (HWP), quarter-wave plate (QWP), and liquid crystal retarder (LCR) control parameters, showing LCR phases around 0 and $\pi$ are challenging to achieve.}
\label{fig:comp_threshold_analysis}
\end{figure}

We also observe that in 2 out of 100 trials, the polarization compensation module fails to return the vertical and diagonal states within the predefined tolerance range. These two events correspond to the longest compensation times observed in Fig.~\ref{fig:comp_speed}(b). In these cases, the compensation algorithm cannot identify a suitable combination of QWP and HWP rotation angles and LCR voltages that satisfies the convergence conditions, shown in more detail in Fig.~\ref{fig:comp_threshold_analysis}. 

\remove{Here, Figure~\ref{fig:comp_threshold_analysis}(a) shows all 100 polarization controller changes used to emulate large, sudden changing fiber birefringence. In all cases, the compensation cycles successfully return the system to a compensated state except for two cases. Blue dots show when the compensator is actively attempting to correct the channel, and red dots show when the compensator deems the reference states to be within the specified threshold.} 
\add{Here, Figure~\ref{fig:comp_threshold_analysis} (a) shows all 100 polarization controller changes used to emulate large, sudden changes in fiber birefringence. Blue triangles and circles show two nonorthogonal polarization measurements for V and D, respectively, before the compensator attempts to correct the channel. The red triangles and dots show when the compensator deems the reference states to be within the specified threshold. In all cases, the compensation cycles successfully return the system to a compensated state except for two cases (the 2 green dots outside the D and 2 green triangles outside the V tolerance circles).}

Recall that we set the tolerance $S_1$ parameter of the (V) polarization reference to be less $<$$-$0.999 and the $S_2$ of the (D) polarization reference to be $<\,$0.95. Figure~\ref{fig:comp_threshold_analysis}(b) shows one of the non-convergent compensation cycles from measurement 21 to 45. At measurement 1, the polarization is changed drastically, but the compensator pulls it back within tolerance by measurement 5. At measurement 20, the polarization is again changed drastically, but the system cannot restore it within tolerance. Instead, it oscillates between within tolerance ($-0.999$) and out of tolerance ($-0.995$), as shown in Fig.~\ref{fig:comp_threshold_analysis}(c). Figure~\ref{fig:comp_threshold_analysis}(d) shows the corresponding compensator angles and phases applied during this non-convergent cycle. The behavior indicates that the LCR calibration has too coarse a resolution, requiring a phase level that is not attainable with the current calibration. When the correct phase is missed, the algorithm drives the LCR toward a $2\pi$ phase shift in the opposite direction; however, this value is also missed, causing the system to push back by $2\pi$ in the other direction. This leads to the observed oscillatory behavior. This issue could likely be mitigated through a higher-resolution calibration and/or environmental stabilization of the LCR, which is known to be sensitive to temperature \cite{schnoor_calibration_2020}. Nevertheless, even when the correct phase is missed, the system still maintains a high level of tolerance ($-0.995$), although it does not reach the desired value of $-0.999$.

\section{Conclusion}
QWN offers a potential quantum networking protocol for the management and control of qubits within and across large-scale quantum networks by leveraging the use of classical header references to monitor quantum signals with minimal disturbance. In this study, we show how QWN headers can be used for real-time polarization monitoring and active birefringence compensation for polarization encoded single- and entangled-photon datagrams transmitted over a 48 km of deployed fiber on the Chicago Loop network. We demonstrate that single photon qubit polarization states can be maintained to within \ang{10} on the Poincar\'{e} sphere to their target polarization. \remove{Two-photon} \add{TPI} visibility for polarization entangled states can be maintained \remove{above 79\%}\add{at its original level} over a 12-hour test period, despite environmental fluctuations on the fiber link as well as sudden large changes to the birefringence we purposely introduce to emulate quantum packet rerouting. We additionally characterize our system's response to environmental disturbances over 44\,hours and show that \remove{while the compensator is active for 32\,hours, the average two-photon interference visibilities are above 96.8\% and 93.8\% in HV and DA bases, respectively. They drift to 95.3\% and 81\% over a 12\,hour period afterward when the compensator is turned off.}\add{,once compensation is disabled at hour 32, the average raw TPI visibilities decline by 0.9 and 7.7\% in the HV and DA bases, respectively.}

For this initial demonstration of QWN channel birefringence compensation, we use motorized waveplates and liquid crystal retarders to reliably recover polarization states in the quantum channel on time scales of $\sim$10\,s. However, compensation speeds can be increased to $\sim$500\,ms using an all-liquid crystal retarder system or to $\sim$100\,ns using electro-optic polarization controllers \cite{chapman_continuous_2024}. Because our system purposely leverages wavelength division multiplexing, we purposely place classical reference headers at a different wavelength channel than the quantum payload, a separation of 122\,GHz (1309.7\,nm for headers and 1309\,nm for the quantum channel). This is slightly more than one channel on the ITU grid at 1310 nm and is promising for practical QWN integration into realistic networks. Polarization mode dispersion, however, limits either the channel spacing or the fiber transmission distance. For further transmission than shown, we would need to reduce the channel spacing. Alternatively, for a larger wavelength channel separation, we would need to reduce the transmission distance of the link. Potential approaches to address this issue include predicting the birefringence evolution over large wavelength spans \cite{eastman_direct_2025}, encoding headers with dual-channel wavelengths \cite{howe_polarization_2025}, or by using ancilla-assisted process tomography \cite{stevens_situ_2025} which offers a larger immunity to PMD. Another possible direction for future work is to investigate whether multi-node QWN configurations could mitigate polarization effects by segmenting long fiber links into shorter sections. In such a scenario, intermediate QWN nodes would measure header polarization states and rewrap the quantum payload before forwarding it to subsequent nodes. However, the feasibility and potential advantages of this approach remain to be analyzed through simulation and experiment.

We also note that future work will need to extend polarization compensation to both signal and idler arms as an important step for practical entanglement distribution and advanced protocols such as teleportation and entanglement swapping over deployed quantum networks. Prior demonstrations incorporate polarization control across multiple links, for example in time-bin entanglement swapping \cite{sun_entanglement_2017} and in configurations where one photon from each source propagates through a deployed fiber while the other is stabilized locally \cite{craddock_high-rate_2026}. These approaches emphasize the need for coordinated polarization control across distributed links. QWN is well-suited for such higher-level control, as classical header references can simultaneously carry network-level information and enable physical-layer stabilization. In particular, header signals associated with signal and idler datagrams could support independent polarization tracking and birefringence compensation across distributed network nodes. Our demonstration, together with these considerations, highlights the potential of QWN as a unified and scalable protocol for integrating network management, control, and physical-layer stabilization in future quantum networks.

\begin{backmatter}
\bmsection{Funding}
This work is supported by the U.S. DOE, Office of Science, under Award Number DE-SC-0022336 and DE-AC02-06CH11357. 

\bmsection{Acknowledgment}
The authors acknowledge the support of StarLight International/National Communications Exchange Facility.

\bmsection{Disclosures}
The authors declare no conflicts of interest.

\bmsection{Data availability} Data underlying the results presented in this paper are not publicly available at this time but may be obtained from the authors upon reasonable request.
\end{backmatter}
\bibliography{refs}

\end{document}